\documentclass[alpha-refs]{wiley-article}

\usepackage{lineno}
\usepackage{afterpage}
\usepackage{siunitx}
\usepackage{hyperref}
\usepackage{float}
\hypersetup{pdfencoding=auto,unicode, psdextra, hidelinks}
\papertype{Submitted to QJRMS}

\title{Meridional Structure and Future Changes of Tropopause Height and Temperature}

\author[1]{Shineng Hu}
\author[2]{Geoffrey K. Vallis}

\affil[1]{Scripps Institution of Oceanography, University of California, San Diego, La Jolla, California}
\affil[2]{Department of Mathematics, University of Exeter, Exeter, EX4 4QF, UK}

\corrauthor{Shineng Hu}
\corraddress{Scripps Institution of Oceanography, University of California, San Diego, La Jolla, California}
\corremail{s3hu@ucsd.edu}

\usepackage[scaled=1.2]{newtxtext}
\usepackage[scaled=1.2, varg]{newtxmath}
\usepackage{xspace}
\newcommand{\TLR}{tropospheric lapse rate\xspace}
\newcommand{\OLR}{outgoing longwave radiation\xspace}
\newcommand{\OLRT}{\text{OLR}}
\newcommand{\TOA}{top of atmosphere\xspace}
\newcommand{\RE}{radiative equilibrium\xspace}
\newcommand{\RDE}{radiative-dynamical equilibrium\xspace}
\newcommand{\WVP}{water vapour path\xspace}
\newcommand{\GMST}{global mean surface temperature\xspace}
\newcommand{\re}{\text{re}}
\newcommand{\rde}{\text{rde}}
\newcommand{\win}{\text{win}}
\newcommand{\lw}{\text{lw}}
\newcommand{\ds}{\text{ds}}
\newcommand{\ws}{\text{ws}}

\newcommand{\pp}[3][]{{\partial^{#1} #2 \over \partial #3^{#1}}}

\newcommand{\figref}[1]{Fig.\ \ref{#1}}
\newcommand{\COT}{CO\textsubscript 2}
\runningauthor{Hu and Vallis}
\begin{document}
\maketitle
\begin{abstract}

\end{abstract}

\section*{Abstract} We use a simple, semi-analytic, column model to better understand the meridional structure of the tropopause height and the future changes in its height and temperature associated with global warming. The model allows us to separate the effects of tropospheric lapse rate (TLR), optical depth, outgoing longwave radiation (OLR) and stratospheric cooling on the tropopause height. When applied locally at each latitudinal band the model predicts the overall meridional structure of the tropopause height, with a tropical tropopause substantially higher than that in higher latitudes and with a sharp transition at the edge of the extratropics. The large optical depth of the tropics, due mainly to the large water vapour path (WVP), is the dominant tropospheric effect producing the higher tropical tropopause, whereas the larger tropical lapse rate actually acts to lower the tropopause height.  The dynamical cooling induced by the stratospheric circulation further lifts the thermal tropopause in the tropics resulting in it being significantly cooler and higher than in mid- and high latitudes. 

The model quantifies the causes of the tropopause height increase with global warming that is robustly found in climate integrations from the fifth Coupled Model Intercomparison Project (CMIP5). The large spread in the increase rate of tropopause height in the CMIP5 models is captured by the simple model, which attributes the dominant contributions to changes in WVP and TLR, with changes in CO$_2$ concentration and OLR having much smaller effects. The CMIP5 models also show a small but robust increase in the tropopause temperature in low latitudes, with a much smaller increase in higher latitudes. We suggest that the tropical increase may at least in part be caused by non-grey effects in the radiative transfer associated with the higher levels of water vapour in the tropics, with near constant tropopause temperatures predicted otherwise.

\section{Introduction}

The tropopause is the boundary separating the relatively quiescent stratosphere and the relatively active troposphere. The thermal stratification in the stratosphere is constrained primarily by radiative processes and,  to a lesser extent, by the slow large-scale overturning circulation. In contrast, the tropospheric lapse rate is maintained by faster dynamical processes acting on much shorter timescales such as moist convection in the tropics and baroclinic eddies in the extratropics \citep[e.g.,][]{manabe1964, held1982, vallis2017}.  The height and temperature of the tropopause control many aspects of the Earth's climate system --- for example the horizontal and vertical scales of the baroclinic eddies and the amount of water vapour in the stratosphere, respectively --- so that understanding its structure, and whether and how that might change in the future, is of fundamental importance. 

\begin{figure} [t!]
\centering{\includegraphics[width=0.6\textwidth]{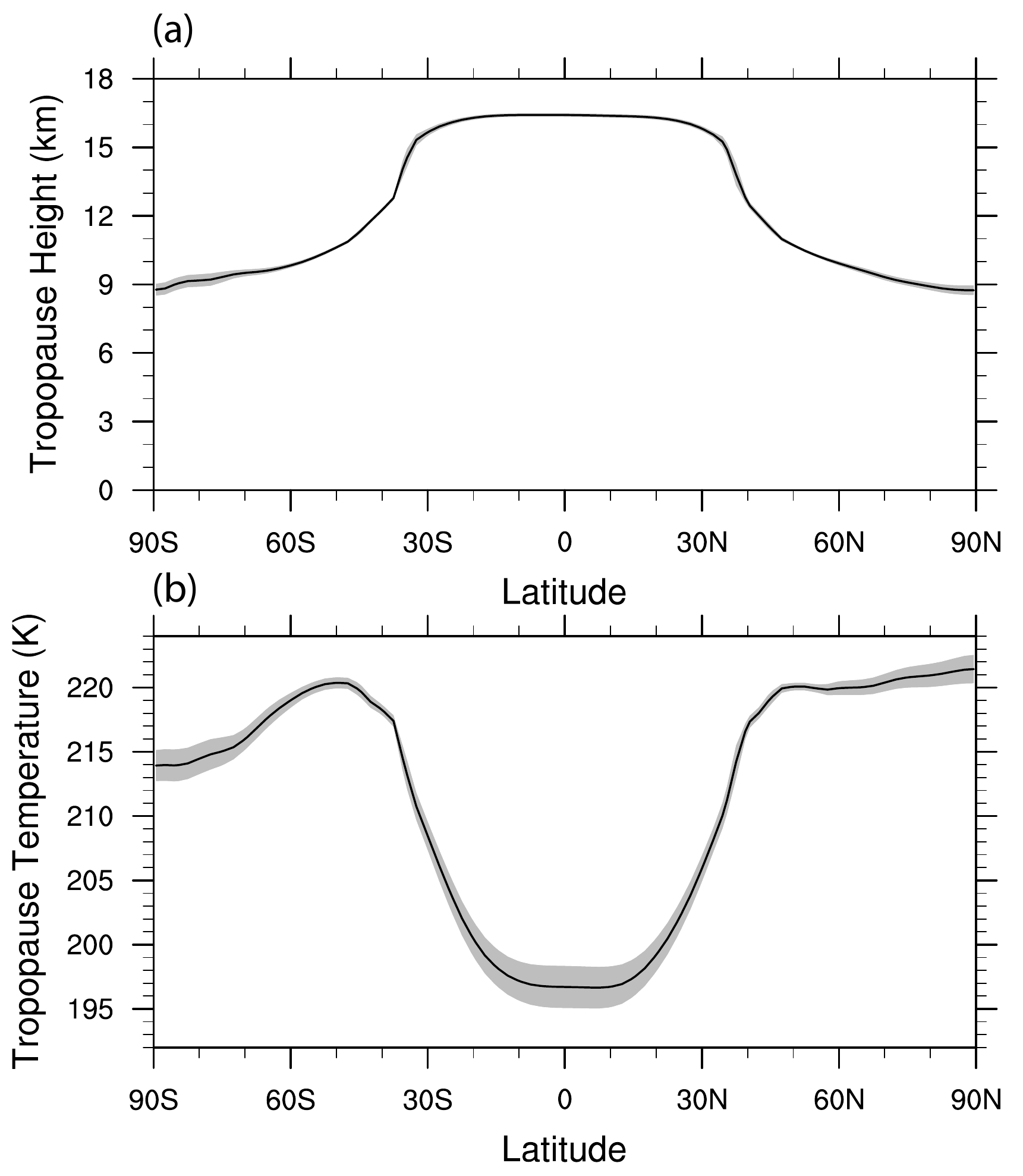}}
\caption{(a) Present-day annual-mean zonal-mean tropopause height averaged over the period of 1979-2017. The grey shading highlights the interannual variations measured by standard deviation. (b) Same as (a) but for the tropopause temperature. Derived from the NCEP2 product.}
\label{obs_trop}
\end{figure}

Various definitions of tropopause height have been proposed based on, for example, lapse rate, potential vorticity, or the level to which the active circulation reaches, and certainly one definition might be more appropriate than another for any given purpose \citep{wmo1957, danielsen1968, holton1995, wilcox2012}. Nevertheless, they all agree on the most prominent feature, namely that the annual-mean zonal-mean tropical tropopause is significantly higher than that over the polar regions (about 16 km versus 8 km) with a fairly sharp transition at the edge of the Hadley Cell as the tropics transitions to the extratropics (Fig. \ref{obs_trop}a). Our goals in this paper are to better understand the overall height and meridional structure of tropopause height, and whether and why the tropopause height and temperature might change in the future. The meridional structure is, in fact, quite well simulated by the current generation of climate models, for example those in the CMIP5 archive. However, simulation is not always the same as understanding and, of perhaps more practical importance, the models in the archive respond in different ways and by different amounts under global warming. Here we will seek to clarify these issues via the construction and use of a relatively simple model that enables us to specifically attribute the meridional structure and temporal changes in tropopause height and temperature to specific changes in lapse rate, optical path and other parameters. 

The basic problem is the determination of the overall height of the tropopause, and a key step in constructing a model for that height is to suppose that the lapse rate is determined dynamically in the troposphere and radiatively above.  Radiative balance must be achieved at the top of the atmosphere, and if the optical thickness is small at the tropopause level, and if the atmosphere is grey in the infra-red, then the temperature of the tropopause can be related to the emitting temperature of the atmosphere, a long-standing and fairly easily-derived result \citep[e.g.,][]{vallis2017}.  
If the \OLR is known, thus giving a `radiative constraint' \citep{held1982, Thuburn_Craig97, thuburn2000},  then we may determine the tropopause height, either by way of a  numerical calculation or an approximate analytic expression \citep{vallis2015}. In a grey approximation the temperature of the tropopause is fixed  (if the \OLR is fixed) and the height of the tropopause then varies inversely with the \TLR (i.e., $-dT/dz$), since the tropopause has to shift downward in order to maintain an unchanged emission temperature if lapse rate increases.  Similarly, if the optical depth increases the tropopause height increases (an extreme example is Venus, where the tropopause is at about 60 km because of the very large greenhouse effect).

Assuming the optical depth is known, the problem is closed if the \TLR is known, and in the tropics and even the subtropics the lapse rate is, to a good approximation, given by the moist adiabatic lapse rate. The mid-latitudes are more complex in this regard,  for lapse rate may be influenced both by moist convection and by baroclinic eddies \citep{juckes2000, schneider2004, zurita2011, zurita2013} and the dynamical constraint, related to the sharpness of the change in lapse rate, may be less tight.  Still, the lower lapse rate of the stratosphere can provide an effective cap to baroclinic instability, and the extratropical tropopause can actually be fairly sharp \citep{birner2006} instantaneously.  In this paper we will take the lapse rate as given, noting that it is actually greater in the tropics than in the mid-latitudes \citep{stone1979, mokhov2006}.  The fact that the tropical tropopause is higher than that in mid-latitudes is then puzzling, because the inverse dependence of tropopause height on lapse rate would suggest a lower tropical tropopause. Evidently, other factors play a role in determining the meridional tropospheric structure, and our first goal is to understand that structure.

Our second goal is to understand the future changes in tropopause height and temperature with global warming. The increase in tropopause height under greenhouse warming is one of the most robust features of simulations with comprehensive climate models \citep{lorenz2007, lu2008, vallis2015}, and is consistent with the observed trend over the recent decades \citep{santer2003}. \citet{vallis2015} argue that the twin causes of the tropopause lifting are the reduction in \TLR (because of a change in the moist adiabatic lapse rate) and the increase in optical depth. Even if this is the case, these factors have not been quantified and there is considerable inter-model spread in the CMIP5 results that is not well understood. Changes in tropopause temperature are also not well understood: grey models with an optically thin stratosphere predict an unchanging tropopause temperature \citep{vallis2015} but CMIP results show tropopause temperatures increasing (although the increase is smaller than that at the surface) and this result is poorly understood.  

In order to better understand and explain all these phenomena, we will develop and use a simple column model with the minimal physics needed.  The advantage of using such a model is that we can explicitly see which factors influence the height and temperature of the tropopause, and how they change with latitude and time.  We begin in Section \ref{sec:data} by briefly describing the main datasets used in this study, including a reanalysis product for the present climate and the CMIP5 model outputs for a future warmer climate. In Section \ref{sec:model} we describe the tropopause model and discuss the key assumptions and the model sensitivity. In Section \ref{sec:meridional} we apply the model to understand the meridional structure of tropopause height in the present climate, and in Section \ref{sec:stratosphere} we discuss the impact of stratospheric circulation. In Section \ref{sec:future} we look at possible changes in tropopause height associated with global warming. In Section \ref{sec:TT}, we extend the tropopause model to have a infra-red window with implications for the future changes in tropopause temperature. We summarize and conclude in Section \ref{sec:conclusions}.

\section{Data and methods} \label{sec:data}

In this paper we will consider only the annual-mean zonal-mean climate states. For the present climate, we use NCEP Reanalysis 2 (hereafter NCEP2) data provided by the NOAA/OAR/ESRL PSD, Boulder, Colorado, USA, from their Web site at https://www.esrl.noaa.gov/psd/. We define the height of the tropopause using the thermal definition provided by the World Meteorological Organization (WMO). For each latitude, we firstly compute the zonal-mean annual-mean temperature and then identify the lowest level where the lapse rate drops below 2\textdegree C/km and remains smaller than 2\textdegree C/km within 2 km above. Next, we interpolate the temperature between the tropopause and the surface onto even height levels, and compute the mean \TLR using a linear regression analysis. The present mean state is defined as the long-term average of monthly data over the period of 1979-2017.

To study the future warming climate, we use the CMIP5 model outputs for the 1pctCO2 scenario. In this scenario, CO$_2$ concentration steadily increases by 1\% per year (denoted as "1\% scenario" hereafter), which corresponds to a quadrupling of CO$_2$ level by year 140. The whole 140-year period of integration is used for linear trend analysis, and the phrase "change" due to greenhouse warming then refers to the trend multiplied by one century. The tropopause height and the \TLR within each climate model are defined in the same way as for NCEP2. We obtain datasets from 24 general circulation models (GCMs) that match the 140-year requirement and contain all variables needed. The GCMs include ACCESS1-0, ACCESS1-3, CanESM2, CCSM4, CMCC-CM, CNRM-CM5, CNRM-CM5-2, CSIRO-Mk3-6-0, CSIRO-Mk3L-1-2, GFDL-CM3, GISS-E2-H, GISS-E2-R, HadGEM2-ES, inmcm4, IPSL-CM5A-LR, IPSL-CM5A-MR, IPSL-CM5B-LR, MIROC5, MIROC-ESM, MPI-ESM-MR, MPI-ESM-P, MRI-CGCM3, NorESM1-M, and NorESM1-ME.

\section{A simple tropopause model} \label{sec:model}

We now describe the essential components of a tropopause model based on the one presented in \citet{vallis2015}, which in turn draws from \citet{held1982} and \citet{thuburn2000}. We assume an atmosphere that is grey in the infra-red, and write the longwave radiation transfer equations \citep[e.g.,][]{goody1964}, as
\begin{equation}
\label{eq:dDdU}
     \frac {\partial D} {\partial \tau} = B - D, \qquad \frac {\partial U} {\partial \tau} = U - B.
\end{equation}
where $D$ and $U$ are downward and upward infra-red irradiance, respectively,  $B=\sigma T^4$ follows the Stefan-Boltzmann law, with $\sigma=5.67\times 10^{-8}$~W~m$^{-2}$. $\tau$ is optical depth increasing downward. At the top of the atmosphere where $\tau=0$, we have the upper boundary conditions $D = 0$ and $U = \OLRT$ (the outgoing longwave radiation), which we assume is given (or taken from observations). 

For the ease of calculation, we define two variables, $I$ and $J$, using the linear combinations of $U$ and $D$, as follows,
\begin{equation}
\label{eq:IJ}
     I = U - D, \qquad J = U + D.
\end{equation}
Eq.~(\ref{eq:dDdU}) can thus be rewritten as, equivalently,
\begin{equation}
\label{eq:dIdJ}
     \frac {\partial I} {\partial \tau} = J - 2B, \qquad \frac {\partial J} {\partial \tau} = I.
\end{equation}
The upper boundary conditions become: $I= \OLRT$ and $J= \OLRT$ at $\tau=0$. 

\begin{figure} [t!]
\centering{\includegraphics[width=0.8\textwidth]{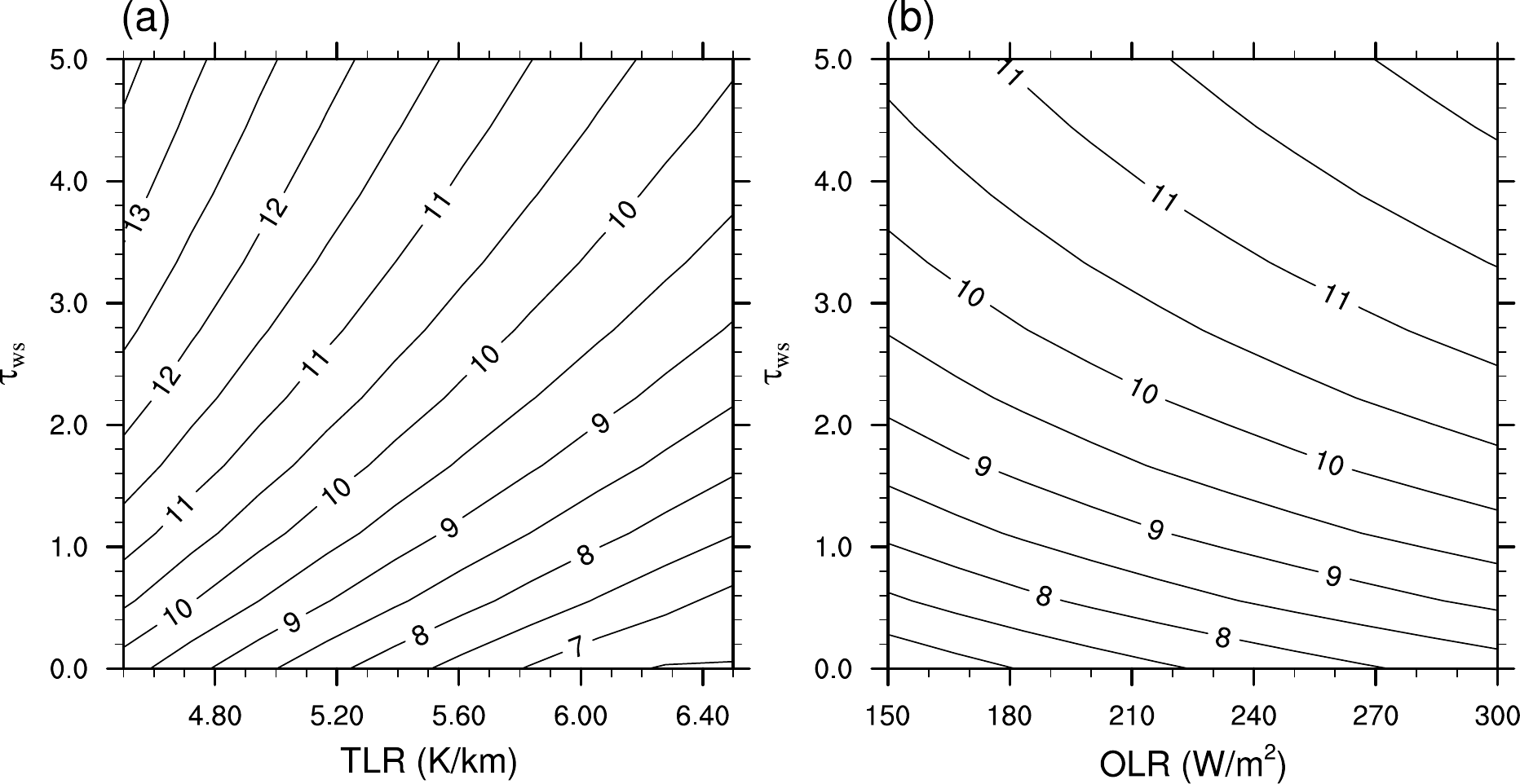}}
\caption{Numerical solutions of tropopause height as a function of $\tau_{\ws}$ and (a) \TLR, and (b) \OLRT. We set a constant value of $\tau_{\ds} = 1$. In panel a, we set OLR = 225 W/m$^2$. In panel b, we set TLR = 5.5 K/km.}
\label{num_sens}
\end{figure}

We assume that the stratosphere is in radiative equilibrium (RE), where the convergence of longwave radiation vanishes, namely,
\begin{equation}
\label{eq:dIzero}
     \frac {\partial I} {\partial \tau} = 0 \qquad \text{and} \qquad \pp I z = 0.
\end{equation}
A consequence of this is that, in \RE, $I$ is constant in the vertical,
\begin{equation}
\label{eq:I}
     I = \OLRT.
\end{equation}
Combining Eq.~(\ref{eq:dIdJ}), Eq.~(\ref{eq:I}), and the \TOA boundary condition $J = \OLRT$, we get
\begin{equation}
\label{eq:J}
     J = (\tau+1)\OLRT.
\end{equation}
Then, based on Eq.~(\ref{eq:IJ}), we can easily derive the expressions of $D$ and $U$,
\begin{equation}
\label{eq:re_soln}
     D,U = \left( \frac{\tau}{2}, \frac{\tau+2}{2} \right) \OLRT.
\end{equation}
and from Eqs.~(\ref{eq:dIdJ}) and (\ref{eq:dIzero}) we see
\begin{equation}
\label{eq:B}
     B = {J \over 2} =  \left(\frac{\tau+1}{2}\right)\OLRT.
\end{equation}
Using the Stefan-Boltzman law, we get the \RE temperature profile for the stratosphere,
\begin{equation}
\label{eq:Tst}
     T_{\re} = \left[ \left(\frac{\tau+1}{2\sigma}\right)\OLRT \right]^{\frac{1}{4}}.
\end{equation}
Note that the \RE solutions above are determined by $\OLRT$ and $\tau$ only and do not depend on the lower boundary conditions on the ground. 

Further, we assume a stratosphere in radiative equilibrium governed by Eq.~(\ref{eq:re_soln}) and a uniformly stratified troposphere, separated by a tropopause at $z=H_T$,
\begin{equation}
\label{eq:Tz}
	T(z) = \begin{cases}
	T_{\re}, & z \geq H_T, \\
	T_T+\Gamma(H_T-z), & H_T \geq z \geq 0,
	\end{cases}
\end{equation}
where $\Gamma=-dT/dz$ is the \TLR and the tropopause temperature $T_T=T_{\re}|_{z = H_T}$. The lower boundary condition at the surface requires that $U=\sigma T_s^4$ at $z=0$, where $T_s$ is the surface temperature (i.e. no ground temperature jump). 

We specify the vertical profile of optical depth as
\begin{equation}
\label{eq:tau1}
	\tau(z) = \tau_{\ws}\exp(-z/H_a)+\tau_{\ds}\exp(-z/H_s),
\end{equation}
where $\tau_{\ws}$ and $\tau_{\ds}$ are the surface optical depths associated with water vapour and dry air, respectively. $H_a = 2$ km and $H_s = 8$km are the scale heights of water vapour and dry air, respectively. We assume that atmospheric pressure exponentially decreases with height $p=p_s\exp(-z/H_s)$, where surface pressure $p_s = 1000$ hPa. Therefore the vertical coordinates of $\tau$, $p$ and $z$ can be converted.

After specifying OLR, $\Gamma$, $\tau_{\ws}$ and $\tau_{\ds}$, we obtain numerical solutions of tropopause height by iterating over the different values of $H_T$ until the lower boundary condition is matched.
(This procedure differs from that in \citet{thuburn2000} since we specify the \OLR, not surface temperature.)   The model sensitivity to the key variables in the ranges that are relevant to the Earth's present climate is shown in Fig.~\ref{num_sens}. When the surface optical depth increases, the emission height increases (to keep the emission temperature constant) and the tropopause also must rise (Fig.~\ref{num_sens}a). When \TLR decreases (as it would in a wetter atmosphere) the tropopause height again must increase to keep the emission temperature constant. Finally, for a fixed lapse rate, an increase in \OLR also leads to a higher tropopause (Fig.~\ref{num_sens}b). This result is not quite obvious since a higher tropopause implies a lower temperature, but a higher \OLR also gives rise to a warmer troposphere as a whole. We note that the direct dependence of tropospheric height on \OLR itself is relatively weak.

\subsection{Approximate analytic solution} \label{sec:analytic}

In some cases the above model admits of an approximate analytic solution \citep{vallis2015}. If the optical depth varies simply as $\tau(z) = \tau_s \exp(-z/H_a)$ and $H_a$ is much less than the height of the tropopause, so that the optical depth is small in the stratosphere, then we find 
\begin{equation}
	\label{trop_analytic} 
    H_T = \frac{1}{16\Gamma}
       \left(\text{C} T_T + \sqrt{\text{C}^2 T_T^2 + 32 \Gamma \tau_s H_a T_T} \right), 
\end{equation}
where $C =  \log 4 \approx 1.4$,  $\Gamma$ is the lapse rate, $T_T$ is the temperature at the tropopause, $\tau_s$ is the surface optical depth and $H_a$ is the scale height of the main infrared absorber. We will not use Eq.~(\ref{trop_analytic}) for calculations in this paper (since more general numerical results are easily obtained) but the equation is instructive in telling us that the tropopause height varies inversely with lapse rate (as expected), and how it increases with surface optical depth, tropopause temperature (and hence OLR), and the scale height of the absorber. Results from Eq.~(\ref{trop_analytic}) are actually quite similar to those shown in \figref{num_sens}, and  reference to Eq.~(\ref{trop_analytic}) is useful in interpreting the results below.  

The most restrictive assumptions of the model are that the lapse rate is constant in the troposphere and that the radiative transfer is grey. We relax the grey assumption in Section \ref{sec:window}. The constant lapse rate assumption is likely to be poorest in polar regions where there may be a low-level inversion.


\section{Meridional structure of tropopause height}  \label{sec:meridional}

We now apply the tropopause model to the different latitudinal bands and explore the meridional structure of tropopause height. For optical depth (Eq.~\ref{eq:tau1}), we assume that a fixed $\tau_{\ds} = 1$ that does not vary with latitude, and a $\tau_{\ws}$ that gradually decreases with latitude to mimic the poleward reduction of \WVP (denoted WVP). In idealized general circulation models, surface optical depth is often assumed to sinusoidally decrease with latitude \citep{frierson2006, ogorman2008}. Here, we simply assume that $\tau_{\ws}$ varies linearly  with \WVP following $\tau_{\ws}=\alpha \text{WVP}$, where $\alpha = 0.1$ mm$^{-1}$. Since \WVP decreases from about 40 mm at the equator to zero at the poles, the surface optical depth decreases from approximately 5 to 1, similar to the meridional profiles used in those idealized general circulation models (for example, in \citet{frierson2006}, surface optical depth decreases from 6 at the equator to 1.5 at the poles). The resultant climate has a similar meridional structure of surface temperature as the observations with a \GMST of 15\textdegree C.  Note that the meridional heat transport of the atmosphere-ocean system is taken into account by the specification of the \OLR as an upper boundary condition to the model. 

\begin{figure} [t!]
\centering{\includegraphics[width=0.6\textwidth]{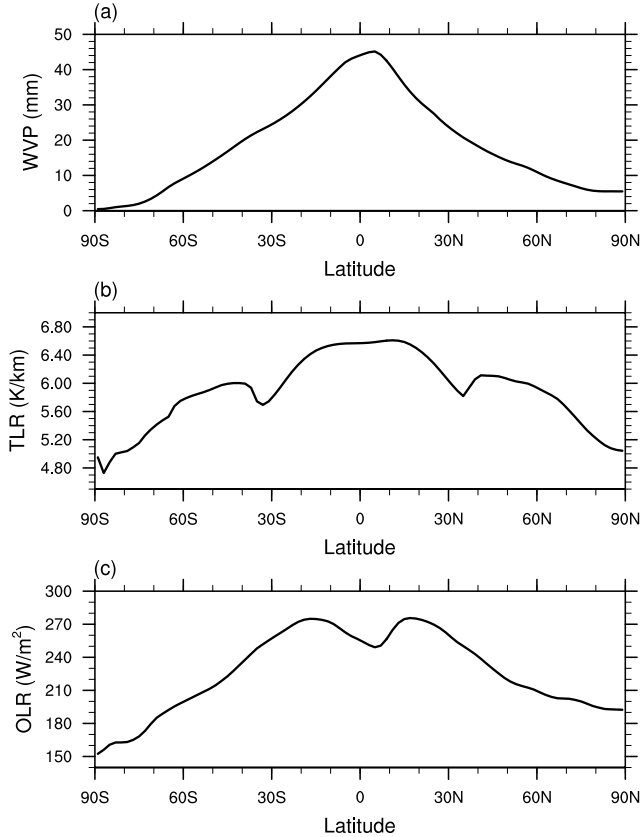}}
\caption{Present-day annual-mean zonal-mean (a) \WVP (WVP) (b) \TLR (TLR) and (c) \OLR (OLR) as a function of latitude averaged over the period of 1979-2017. Derived from the NCEP2 product.}
\label{input}
\end{figure}

The optical path $\tau_s$ (and so water vapour content integrated over the depth of the atmosphere), \TLR (i.e., $\Gamma$) and \OLR are the three key variables in our model (as can be seen from Eq.~\ref{trop_analytic} in which $T_T$ parameterizes the OLR), and they are all characterized by substantial meridional variations. Fig.~\ref{input} shows their annual-mean zonal-mean meridional profiles computed from the NCEP2 product. Water vapor content peaks at the equator with a maximum value of 40 mm, decreases almost linearly with latitude, and nearly vanishes over the polar regions (Fig.~\ref{input}a). The outgoing longwave radiation decreases poleward as the layers emitting infra-red radiation get cooler, and the equator-to-pole contrast is about 100 W/m$^2$ (Fig.~\ref{input}c).

In the tropical atmosphere, horizontal tropospheric temperature gradients are relatively small within 30\textdegree S--30\textdegree N, while surface temperature is uniform only in a narrower equatorial band within 15\textdegree S--15\textdegree N. As a result, \TLR experiences non-monotonic changes with latitude: it reaches the highest value of 6.5 K/km following the moist adiabats in the equatorial band, starts to decrease at 15\textdegree of latitude, begins to increase at 35\textdegree of latitude, and finally decreases towards the poles where the lowest value of 5 K/km occurs (Fig.~\ref{input}b).

\begin{figure} [t!]
\centering{\includegraphics[width=0.6\textwidth]{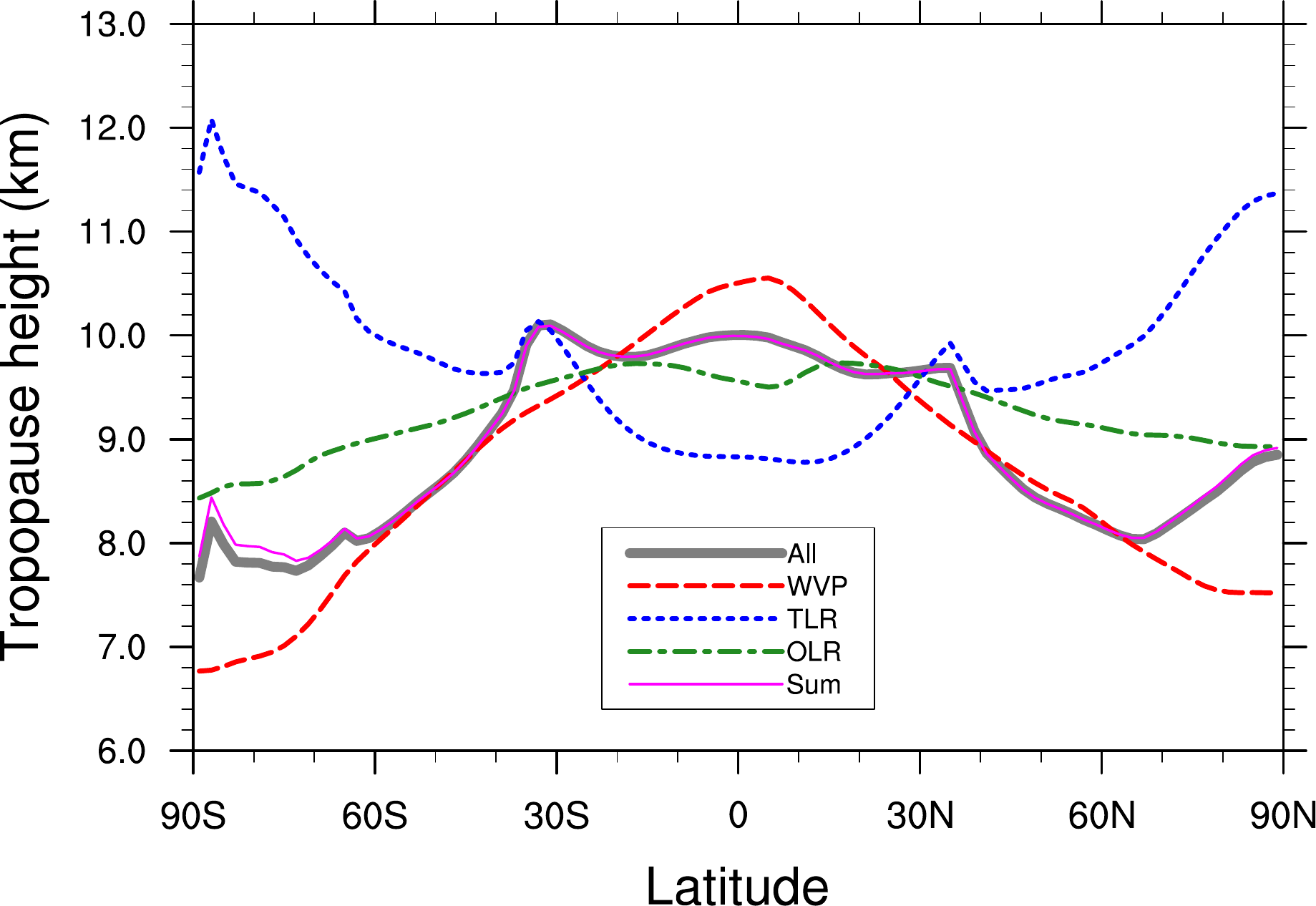}}
\caption{The model-predicted tropopause height as a function of latitude using input information from NCEP2. The red line corresponds to the model-predicted tropopause height with only the latitude dependence of \WVP (WVP) while keeping the other two factors at their global mean values. Similarly, the blue and green lines are for model-predicted tropopause height with only the latitude dependence of \TLR (TLR) or \OLR (OLR), respectively. The thick grey line is for the tropopause height with the latitude dependence of all three factors included, and the thin magenta line is the linear sum of the  three curves,  with an offset for better comparison with the thick grey curve.}
\label{pred_trop}
\end{figure}

Next we use those three meridional profiles (i.e., \WVP, \TLR and \OLR) as inputs for our tropopause model to predict the meridional structure of tropopause height (Fig.~\ref{pred_trop}). To isolate their individual contribution, we conduct three sensitivity cases where we only allow one of them to vary with latitude while fixing the other two at their global mean values for all latitudes. As noted earlier, \TLR alone leads to a lower tropopause in the tropics compared to the polar tropopause, which is not in accord with the observations. Nevertheless, the greater \WVP (i.e. thicker optical depth) in the tropics instead results in a much higher tropopause than that over the poles by about 4 km. The impact of \OLR on the tropopause height is relatively small; a 100 W/m$^2$ equator-to-pole contrast of \OLR only leads to 1 km difference in the tropopause height. 

With all the three factors included, the tropopause model produces a reasoably realistic meridional structure of tropopause height (Fig.~\ref{pred_trop}). The competing effects of \WVP, \TLR and \OLR lead to an almost flat tropopause in the tropics. The sharp transition of tropopause height occurs at about 35\textdegree  S (35\textdegree  N), which results from the non-monotonic change of \TLR near the edge of the tropics (Fig.~\ref{input}b). Outside the tropics, the tropopause height generally decreases with latitude except for the slight reversal poleward of 65\textdegree N. All those features resemble that in observations. Still, the equator-to-pole contrast in tropopause height predicted by the model (2 km) is too small as compared with the observations (8 km). In the next section, we will revise the simple model to include the tropical dynamical cooling effects of the stratospheric circulation to account for such discrepancies.

 \section{The role of stratospheric circulation} \label{sec:stratosphere}
 \begin{figure} [t!]
 \centering{\includegraphics[width=0.9\textwidth]{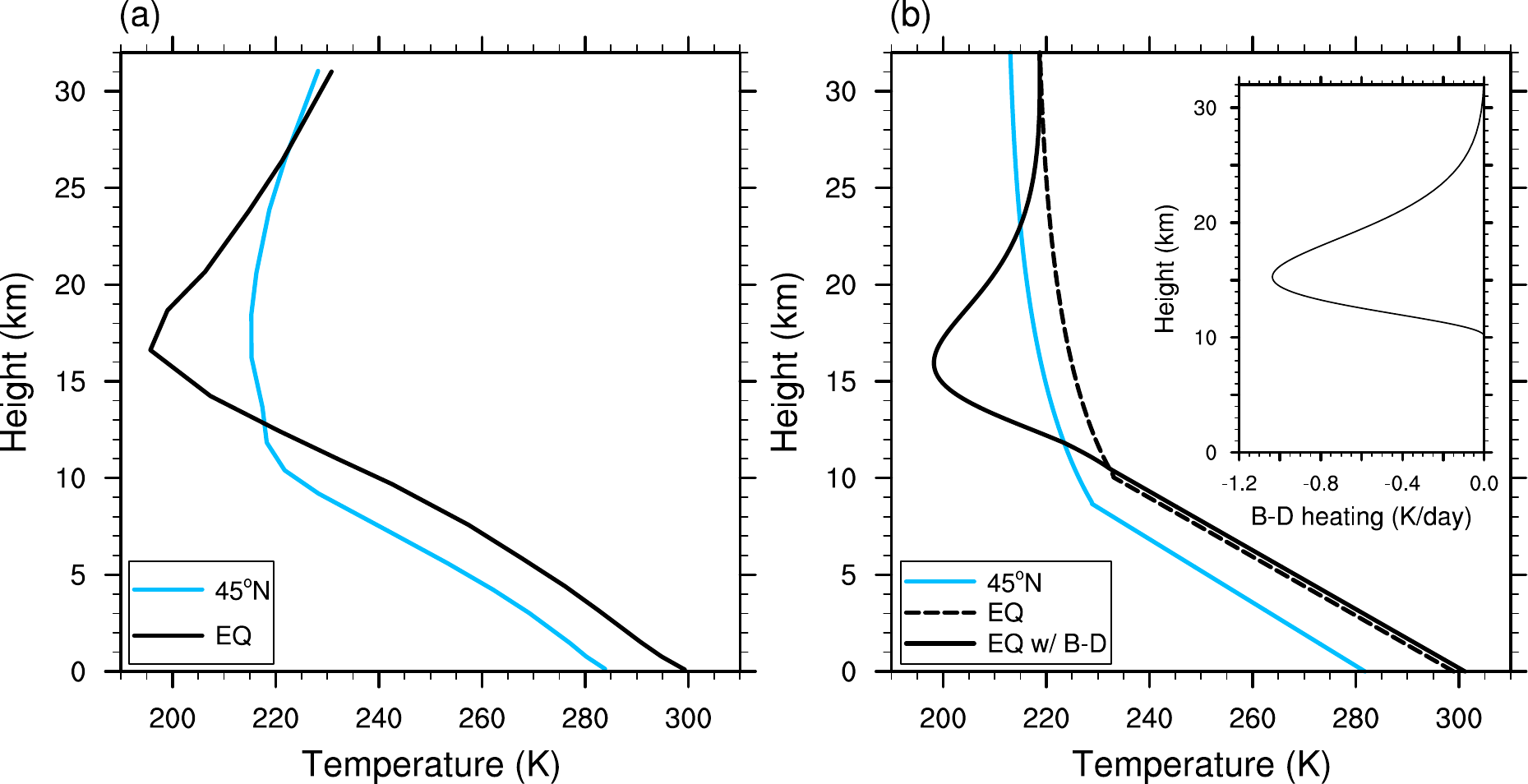}}
 \caption{(a) Annual-mean zonal-mean temperature profiles at the equator (black) and 45\textdegree N (light blue) in NCEP2. (b) Same as panel a but derived from the radiative-convective model; the black dashed and solid lines are for equatorial temperature profiles without and with a dynamical cooling representing that of the Brewer-Dobson circulation. The dynamical cooling profile is embedded in panel (b), and its peak value is 1 K/day, similar to that for the current climate.}
 \label{bd_impact}
 \end{figure}
 
As noted, the equator-to-pole contrast of tropopause height produced by the model is too small compared to that seen in observations. Moreover, the tropical tropopause is actually significantly colder than mid-latitude and polar tropopause by about 20 K (Fig.~\ref{obs_trop}b), and this feature is not captured by the model (Fig.~\ref{bd_impact}). Previous studies suggest that the stratospheric circulation that consists of an ascending motion in the tropics and a descent flow over the polar regions, helps shape the meridional structure of tropopause height \citep{thuburn2000, birner2010, haqq2011, zurita2013}. The tropical upward motion in the highly stratified stratosphere induces strong dynamical cooling $Q_s$ that breaks the radiative balance we previously assumed.

As a result, Eq.~(\ref{eq:dIzero}) needs to be modified to become
 \begin{equation}
 \label{eq:dIQs}
      \pp I \tau +Q_s = 0,
 \end{equation}
where $Q_s = \rho c_{p} \dot{Q} d\tau/dz$. Here $\dot{Q}$ is the dynamical heating rate in K/s,  $\rho$ is atmospheric density, and $c_p=1005$ J/(kg$\cdot$K) is the heat capacity of air.
If the vertical profile of $Q_s$ is given, we can (see Appendix) obtain the stratospheric \RDE temperature profile 
\begin{equation}
\label{eq:Tst1}
     T_{\rde} = \left[ \left(\frac{\tau+1}{2\sigma}\right) \OLRT + \frac{Q_s - \overline{\overline{Q_s}}}{2\sigma} \right]^{\frac{1}{4}},
\end{equation}
where $\overline{\overline{Q_s}}(\tau)=\int_{0}^{\tau} \int_{0}^{\tau'} Q_s(\tau'') d\tau'' d\tau'$. After replacing $T_{\re}$ with $T_{\rde}$ in Eq.~(\ref{eq:Tz}), again we can numerically solve the system by iterating over the different values of $H_T$ until the surface boundary condition ($U=\sigma T_s^4$) is matched. (Details of how the numerical solutions are obtained are presented in the Appendix.)   The numerical solution of $H_T$ then gives the boundary between the lower region where the lapse rate is specified and the upper radiative--dynamical region. This may be the top of the convective region but tropopause itself, as given by  the usual WMO lapse-rate definition, may be much higher, as we now discuss.

An example calculation is shown in \figref{bd_impact}b. We impose an equatorial dynamical cooling that is centered around 150 hPa and sinusoidally decreases with pressure upward and downward within 20--280 hPa; its vertical profile is embedded in Fig.~\ref{bd_impact}b. We choose a peak cooling magnitude of about 1 K/day, similar to that found in reanalysis products \citep{fueglistaler2009}. For other variables (i.e. OLR, $\Gamma$, $\tau_{\ws}$ and $\tau_{\ds}$), we use their observed values at the equator that are used to produce Fig.~\ref{pred_trop}. For comparison, we also plot the modeled temperature profile at 45\textdegree N using the corresponding input values as observed, with no Brewer--Dobson cooling.   The imposed dynamical cooling substantially cools the lower stratosphere/upper troposphere. It gives rise to an elevated tropopause layer (16 km based on the cold point) that is significantly cooler than the mid-latitude tropopause, in agreement with observations (Fig.~\ref{bd_impact}a) and broadly consistent with \citet{thuburn2000}. The tropopause height here is distinct from, and generally above, the location of $H_T$ (which is at 10.5 km), and is suggestive of a tropopause (or at least a thermal tropopause, such as given by the WMO definition) that sits above the top of the convective region.  Finally,  note that our model does not show as large a rise in temperature above the tropopause as is seen in the observations because of the absence of ozone heating, but this could easily be added in.

\section{Tropopause Height Changes in a Warmer Climate} 
\label{sec:future}

\subsection{CMIP5 analyses}

\begin{figure} [t!]
\centering{\includegraphics[width=0.58\textwidth]{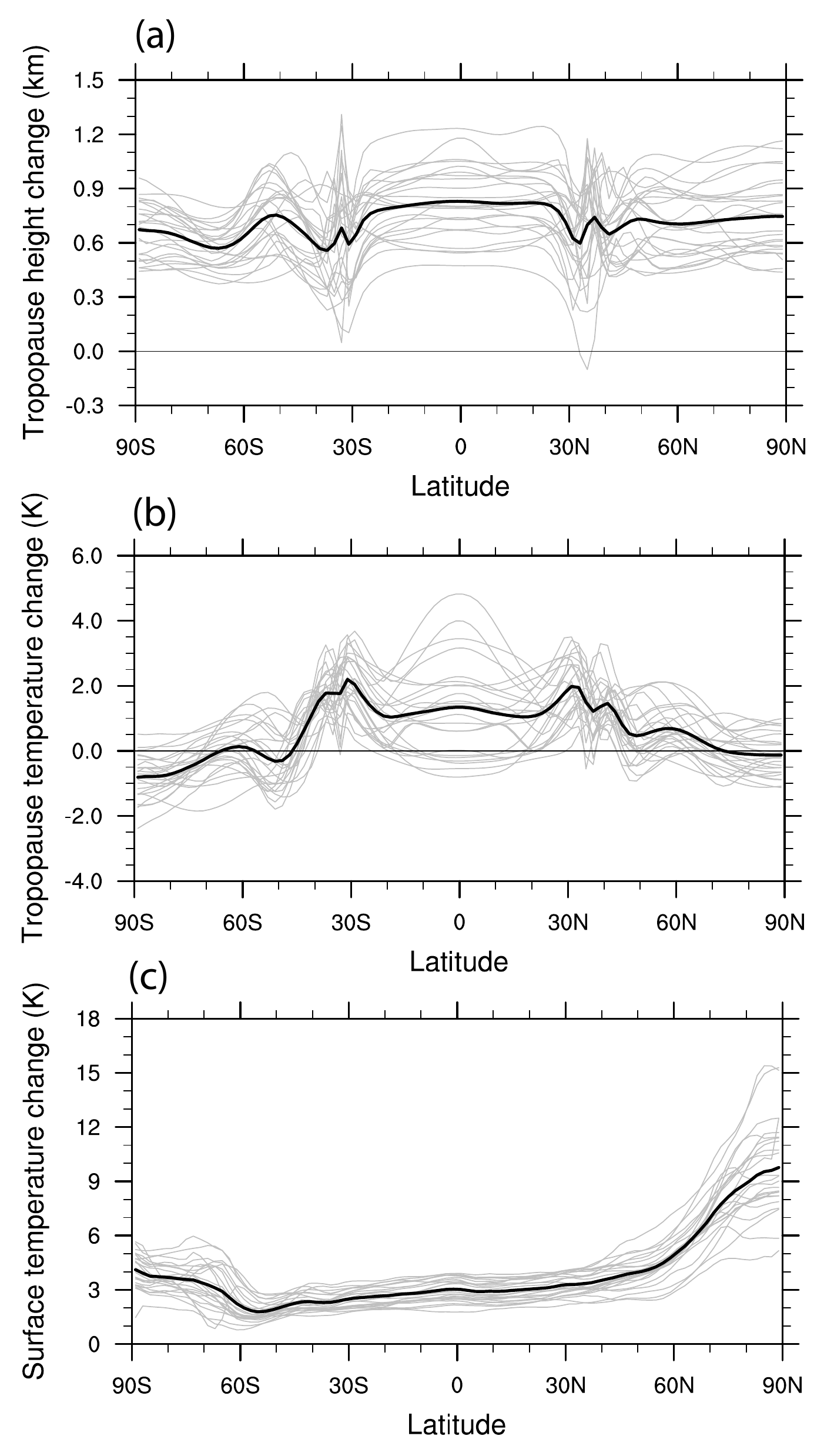}}
\caption{Changes over a century in annual-mean zonal-mean (a) tropopause height, (b) tropopause temperature and (c) surface temperature for the 1\% scenario as a function of latitude. Grey lines are for each individual CMIP5 model, and black lines highlight the multi-model averages.}
\label{HT}
\end{figure}

One of the most robust features of atmospheric large-scale circulation response to global warming is the increase of tropopause height \citep{santer2003, lu2008, lorenz2007, vallis2015}. Consistent with these studies, we find a robust upward shift of the tropopause at all latitudes across all the models (Fig.~\ref{HT}a). In the multi-model mean, the tropopause shifts upward with a rate of 0.6-0.8 km per century depending on latitude. Although the increase is robust there is a large inter-model spread in its magnitude, with the spread being particularly large in the tropics. 

\begin{figure} [t!]
\centering{\includegraphics[width=0.6\textwidth]{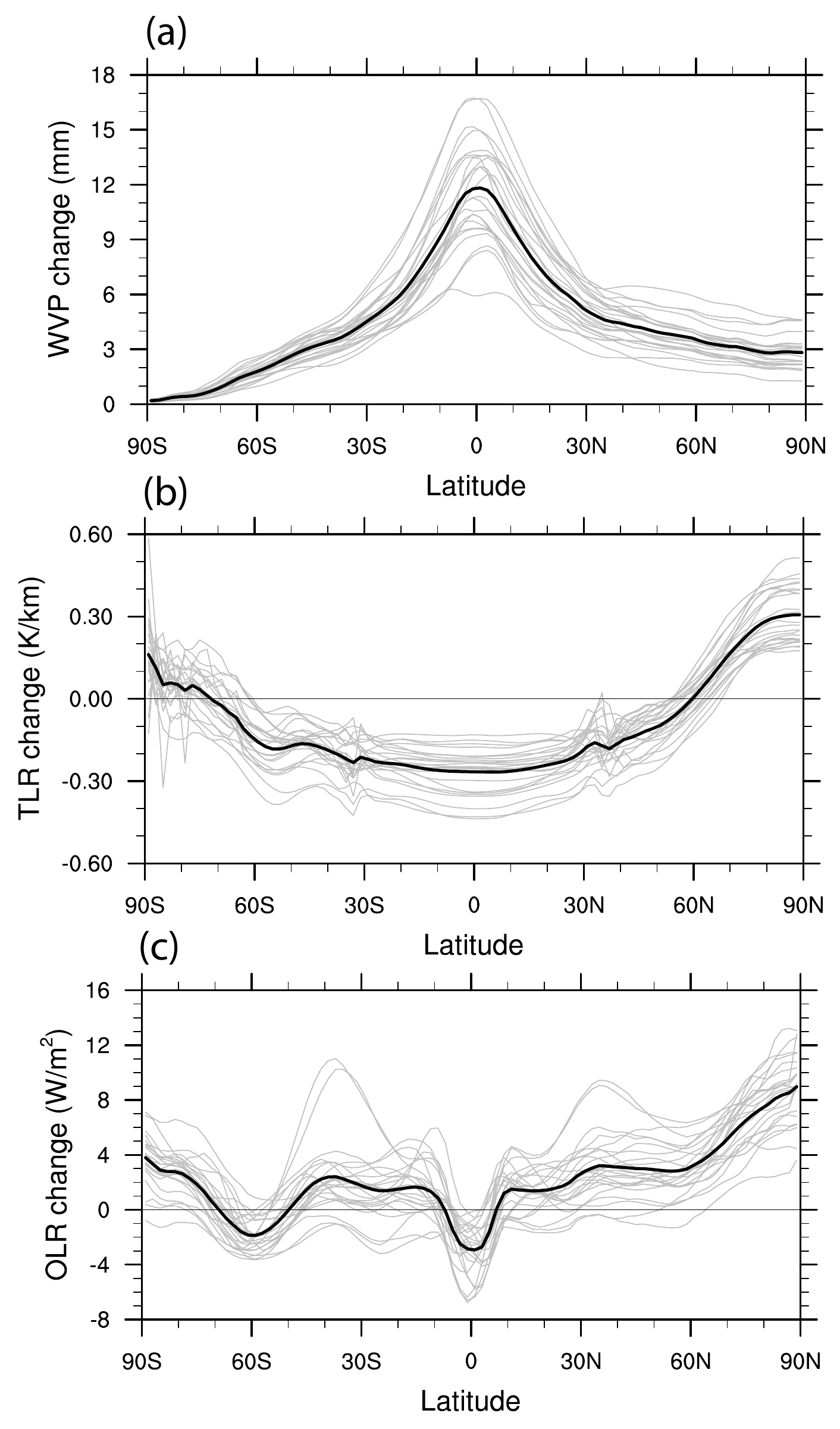}}
\caption{Changes in (a) water vapour path (WVP), (b) tropospheric lapse rate (TLR) and (c) outgoing longwave radiation (OLR) over a century for the 1\% scenario, as a function of latitude. Grey lines are for each individual CMIP5 model, and black lines highlight the multi-model mean. }
\label{cmip_trend}
\end{figure}

\begin{figure} [t!]
\centering{\includegraphics[width=0.6\textwidth]{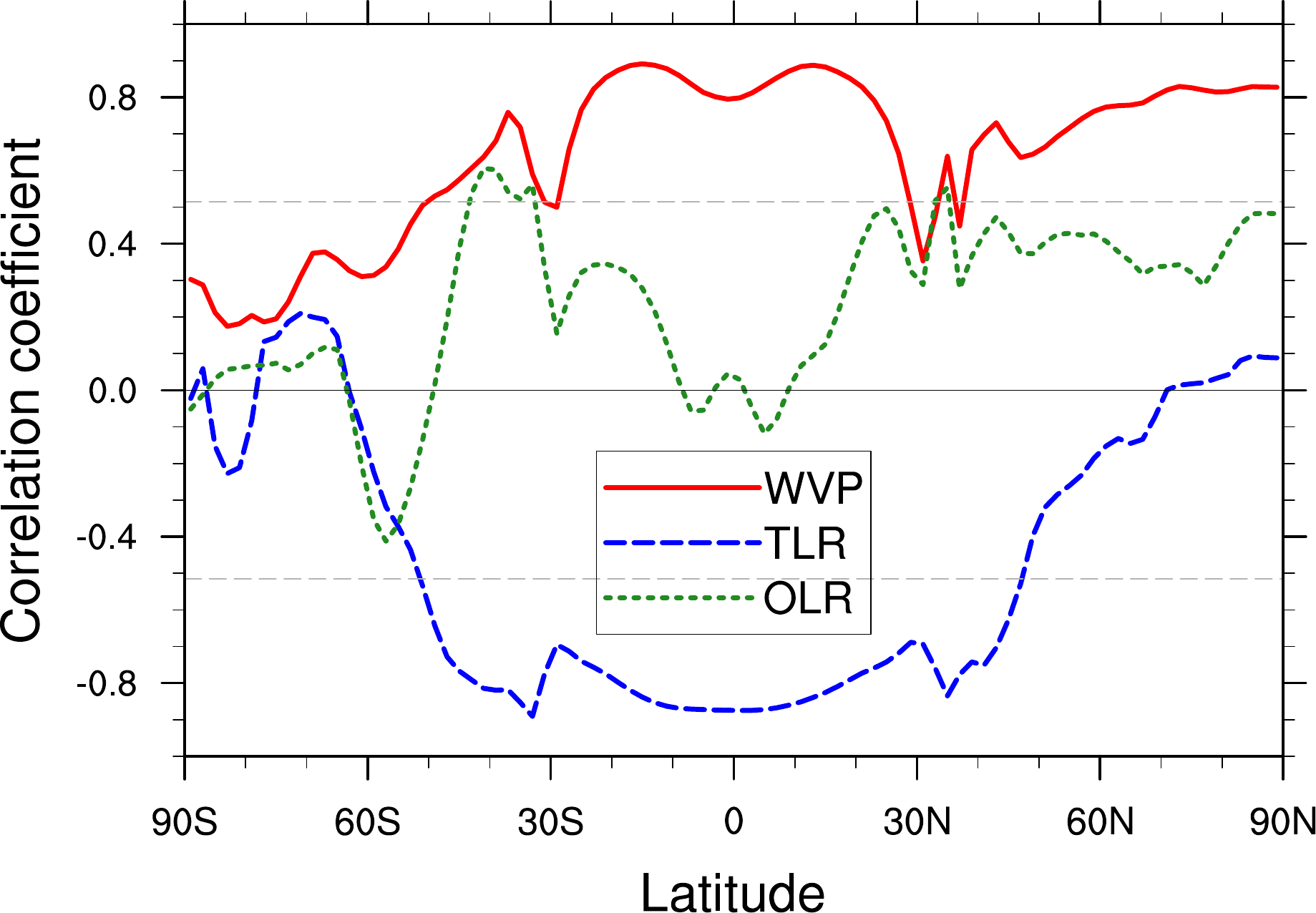}}
\caption{Linear correlation coefficients between the tropopause height changes and the changes in water vapour path (WVP, red solid), tropospheric lapse rate (TLR, blue dashed), or outgoing longwave radiation (OLR, green dotted) among the CMIP5 models, as a function of latitude. The two grey dashed lines highlight the significance level ($p<0.01$) of the Pearson correlation.}
\label{inter_spread}
\end{figure}

To better understand these features we investigate the connection between the changes in tropopause height and changes in \WVP, \TLR and \OLR, the three main controlling factors suggested by our tropopause model. On average, \WVP increases by 12 mm over a century at the equator (Fig.~\ref{cmip_trend}a); its relative change is about 30\% globally and peaks over the high latitudes (40-50\%). The inter-model spread in the \WVP changes are positively and significantly correlated with the tropopause height changes for almost all the latitudes ($p<0.01$) except for the high-latitudes of the Southern Hemisphere, and the highest correlation ($>$0.8) is found over the tropics and the Arctic (Fig.~\ref{inter_spread}a).

The tropospheric lapse rate decreases in the tropics, following the decrease in the saturated adiabatic lapse rate in a warming climate (Fig.~\ref{cmip_trend}b). However, the lapse rate increases over the polar regions, particularly over the Arctic, because polar amplification is mostly confined in the thermally-stable lower troposphere. The inter-model spread of the \TLR changes are negatively correlated with the tropopause height changes over the broad tropics and subtropics within $50^\circ$S--50\textdegree N, and the peak correlation exceeds 0.8 (Fig.~\ref{inter_spread}b).

The \OLR change shows a more complicated meridional structure, and the inter-model spread is larger than the multi-model mean over latitudes (Fig.~\ref{cmip_trend}c). This  structure reflects the fact that \OLR is affected by a variety of factors involving both external forcing (CO$_2$) and internal climate feedbacks (water vapour, clouds), as well as changes in meridional heat transport. The magnitude of the change in the resultant \OLR change is quite small. The inter-model spread of the \OLR changes is only weakly correlated with the tropopause height changes over the Northern subtropics and high latitudes and some of the Southern subtropics, and most of the correlations do not pass the 1\% significance level (Fig.~\ref{inter_spread}c).  Although the net \OLR is positive, the change in top-of-atmosphere net radiation, incoming net shortwave minus outgoing longwave, is actually positive because the incoming shotwave increases, at least partly because of a reduction in ice cover.

\subsection{Causes of tropopause height changes} \label{sec:trop-height}

We now employ our model to quantitatively interpret the tropopause height changes seen in the CMIP5 models. We start from the background meridional structure of tropopause height calculated according to NCEP2 (i.e. the thick grey line in Fig.~\ref{pred_trop}). We perturb the NCEP2 input variables (i.e. \WVP, \TLR and \OLR) with the projected changes in each CMIP5 model, and estimate the resultant changes in the tropopause height. We repeat the procedure for each variable, with the rest being unchanged, to isolate their individual contribution.  

For the impact of CO$_2$, we use a slightly different approach, since the radiative forcing of CO$_2$ is relatively better constrained in climate models. The 1\%/year increase of CO$_2$ will lead to 2.7 times of CO$_2$ concentration after a century, and it corresponds to a radiative forcing of +5.3 W/m$^2$ (scaled from +3.7 W/m$^2$ per CO$_2$ doubling). Based on our model estimate, a reduction of \OLR by 5.3 W/m$^2$ without changes in temperature profiles would require an increase of dry-air surface optical depth $\tau_{\ds}$ in Eq. (\ref{eq:tau1}) by 0.11. Therefore, the impact of the CO$_2$ increase over a century can be estimated by increasing $\tau_{\ds}$ by 0.11 uniformly for each latitude and calculating the changes in the tropopause height. Unlike the aforementioned three factors, we have estimated the effect of CO$_2$ only once, because its uncertainty among the models is relatively small.

\begin{table}[t]
\centering
\begin{tabular}{r c c c c c c}
\hline
\hiderowcolors
& WVP  & TLR & OLR & CO$_2$ & All & CMIP5  \\
\hline
Tropical mean (km) & 0.39$\pm$0.08 & 0.35$\pm$0.11 & 0.00$\pm$0.01 & 0.03 & 0.78$\pm$0.20 & 0.80$\pm$0.18 \\
Global mean (km) & 0.34$\pm$0.07 & 0.26$\pm$0.10 & 0.01$\pm$0.01 & 0.04 & 0.65$\pm$0.17 & 0.74$\pm$0.16 \\
\hline
\\
\end{tabular}
\caption{The CMIP5-informed simple model estimates of tropopause height changes in a warming climate. Tropical-mean and global-mean results are shown separately. Attribution analysis is conducted for each individual factor, including water vapour path (WVP), tropospheric lapse rate (TLR), outgoing longwave radiation (OLR) and carbon dioxide (CO$_2$). `All' refers to the case incorporating all the four factors. `CMIP5' refers to the results computed from the outputs of the 24 CMIP5 model. Both the multi-model mean and inter-model spread, measured by standard deviation, are shown. See the main text for the details of the calculations.}
\label{decomp}
\end{table}

For all the above effects we proceed numerically, since our numerical solutions are more general than the analytic ones,  but the changes may largely  be interpreted by an analysis of Eq.~(\ref{trop_analytic}). 
We conduct the calculations separately for each latitude and for each model, and the multi-model-mean results are summarized in Table \ref{decomp}. With all the four factors considered, the simple model predicts the tropical-mean tropopause height change seen in the CMIP5 models (0.78 km versus 0.80 km) with an error of less than 3\%. Among them, the increase of \WVP acts as the dominant factor (about 48\%), followed by the reduction of \TLR (44\%). Although changes in CO$_2$ is the only external forcing, its direct contribution to the tropopause height increase is quite small (4\%). The effect of changes in \OLR is even smaller, less than 1\%.  These percentages should not be taken as exact, but we quote them because they are indicative of the magnitude of the effects. 

\begin{figure} [t!]
\centering{\includegraphics[width=0.5\textwidth]{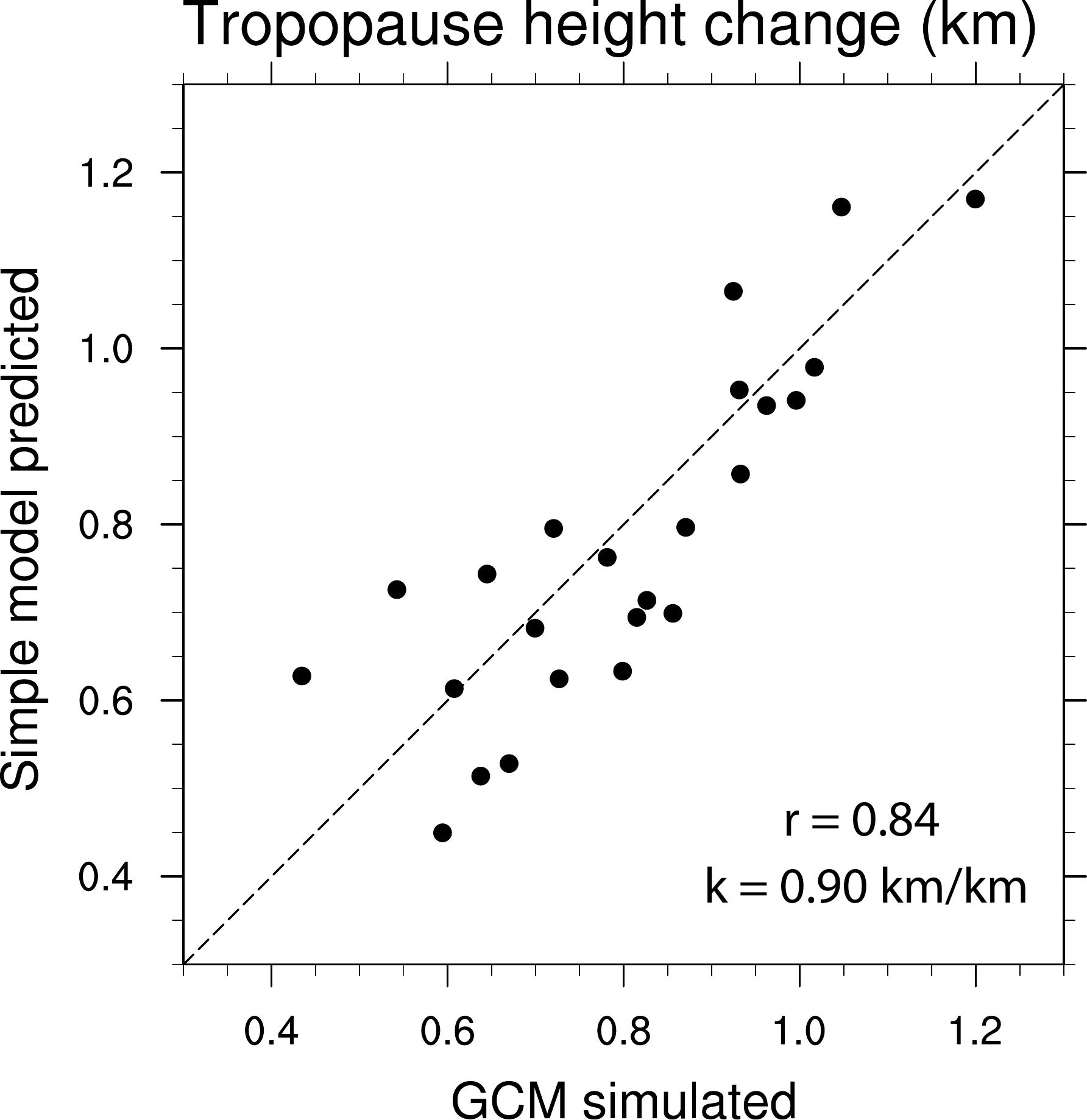}}
\caption{Inter-model scatter diagram for model-predicted versus GCM-simulated tropopause height changes for the 1\% scenario. The correlation ($r$) and regression ($k$) coefficients are shown in the bottom-right corner of the diagram.}
\label{pred_trop_change}
\end{figure}

The model also captures the spread of the tropopause height changes among the CMIP5 models quite accurately (Fig.~\ref{pred_trop_change}), and the inter-model correlation coefficient is as high as $r = 0.84$. The  scattered dots almost fall onto a one-to-one line, suggesting that the simple model predicts the tropopause height change for each individual CMIP5 model quite well. The skill of the simple model is less in the mid-latitudes and breaks down in the high-latitudes (not shown),  consistent with the fact that high correlations are mainly found in the tropics (Fig. \ref{inter_spread}). The lower skill in mid- and high latitudes may be partly due to the lack of `tightness' of the radiative constraint, meaning that small changes in the tropopause height do not necessarily lead to large changes in outgoing radiation \citep{zurita2013}. At high latitudes the presence of a low-level inversion also make the model less appropriate. When averaged over the whole globe, the simple model underestimates the tropopause height changes (Table \ref{decomp}) and the inter-model correlation becomes slightly weaker ($r = 0.82$).    As we noted in Section \ref{sec:analytic} another potential weakness of the model is that it is grey in the infra-red and we look at this effect in Section \ref{sec:twoband}.

\section{Tropopause temperature changes} \label{sec:TT}

We now look at possible changes in tropopause temperature under global warming. Among other things, the tropical tropopause temperature is important in determining the stratospheric water vapour amount \citep{mote1996} and the intensity of tropical cyclones \citep{emanuel2013,  wang2014}.  Now, the emission temperature at any particular latitude will not change significantly unless there are large changes in meridional heat transport, which seems unlikely in the foreseeable future under most warming scenarios.   Under these circumstances, the tropopause temperature should stay roughly constant under greenhouse warming, since this is tied to the emission temperature.  The constant-temperature result is exact in a grey model with an troposphere with a specified lapse rate (e.g., in radiative-convective equilibrium)  and an optically thin stratosphere in radiative equilibrium~\citep{vallis2015}. 

A possibly-related hypothesis is that  the tropical anvil clouds occur at a nearly constant temperature in a warming climate \citep{hartmann2002, kuang2007,  li2012}. This idea, often termed the Fixed Anvil Temperature (FAT) hypothesis, has also been applied to the extratropics \citep{thompson2017} where the clear-sky diabatic mass flux is presumed to vanish at roughly a fixed water vapour concentration, as in the tropics. Although there is no obvious guarantee that the two (tropopause temperature and anvil cloud top) should covary with each other, some observational products indeed assign the cloud-top temperature based on the tropopause temperature \citep{marchand2010}. 

Changes in tropopause temperature (using the WMO definition) under global warming at each latitude in the CMIP5 models are shown in  Fig.~\ref{HT}b. We see that the tropopause temperature in the extratropics and the polar regions (i.e. poleward of 45\textdegree N and 45\textdegree S) indeed varies little compared to the surface warming,  consistent with the radiative and thermodynamic constraints. However, some tropopause warming is found at the tropical tropopause within 30\textdegree S-30\textdegree N, although there is a large inter-model spread, and the warming is only half that at the surface (Fig.~\ref{HT}c).  A similar warming trend at the tropical tropopause has been identified in observations and modelled future warming scenarios \citep{austin2008, kim2013, lin2017},  although a near-constant tropical tropopause temperature was apparently found with a cloud-resolving model by \citet{seeley2019}.  These results do not speak directly to the FAT hypothesis, since the tropical tropopause may be well above the anvil-cloud top and the changes in the two are not necessarily coupled. 

To better understand the causes of the increase in tropopause temperature,  we impose the changes in CO$_2$, \WVP, \TLR and \OLR under the 1\% scenario from each model (as was done in Section \ref{sec:trop-height}). Note that although the model is grey in the infra-red (a restriction we relax below) the stratosphere has a finite optical depth because of the presence of carbon dioxide and is not isothermal, so that the fixed tropopause temperature result is not exact. In fact, as the \WVP increases or the \TLR decreases, the tropopause height will increase and thus the tropopause temperature will actually decrease slightly, if there is an unchanged \RE temperature profile in the stratosphere, given that $T_{\re}=[(\tau+1)\OLRT/(2\sigma)]^{1/4}$.  However, this is not the only effect. As the CO$_2$ concentration increases, its scale height is sufficiently large so that the stratospheric optical thickness increases and thus $T_{\re}$ increases and this leads to an increase in the tropopause temperature. Finally, an \OLR increase (decrease) will tend to increase (decrease) the tropopause temperature by increasing (decreasing) $T_{\re}$. (A rough estimate of the magnitude of the tropopause warming, $\Delta T$ due only to changes in \OLR of $\Delta \text{OLR}$ is obtained by using Eqs.~(\ref{eq:B}) and (\ref{eq:Tst}) with $\tau = 0$. For small $\Delta T$ we have $4\Delta T_T/ T_T = \Delta \OLRT/\OLRT$, where $T_T = T_\re$ is the tropopause temperature, which gives $\Delta T \approx 0.5$\textdegree C.) 

After taking into account all the above factors, the grey model predicts an almost unchanged tropical tropopause temperature (0.03\textdegree C per century), in contrast to the more substantial tropopause warming found in the CMIP5 models (about 1\textdegree C per century on average, shown in \figref{HT}). To better understand this result we will increase the complexity of our radiative model, but in the most minimal way, as follows.

\subsection{An infra-red window} \label{sec:window}
The real atmosphere is not grey to infra-red radiation and atmospheric opacity varies with wavelength, as was known to \citet{Arrhenius96} and \citet{simpson1928}. To capture the essence of the non-greyness, let us construct a model with two bands in the infra-red, a window band (8-\SI{13}{\micro\meter}; denoted with the superscript `win') and the non-window infra-red band (marked with the superscript `lw') where most of the infra-red absorption occurs. (When we refer to the `two-band' model and the 'grey' model we mean two bands in the infra-red, and grey in the infra-red, respectively, without reference to solar radiation.)  The modified radiative transfer equations become
\begin{equation}
\label{eq:dDdUlw}
     \frac {\partial D^{\lw}} {\partial \tau^{\lw}} = \beta B - D^{\lw}, \qquad \frac {\partial U^{\lw}} {\partial \tau^{\lw}} = U^{\lw} - \beta B,
\end{equation}
and
\begin{equation}
\label{eq:dDdUwin}
     \frac {\partial D^{\win}} {\partial \tau^{\win}} = (1-\beta) B - D^{\win}, \qquad \frac {\partial U^{\win}} {\partial \tau^{\win}} = U^{\win} - (1-\beta) B.
\end{equation}
$\beta$ and $1-\beta$ represent the non-window fraction and the window fraction of the emitted infra-red radiation, respectively. This is similar to the model in \citet{weaver1995}, but here we assume a small but finite optical depth for the window region, as in \citet{Geen_etal16} and \citet{vallis2018}, and following those authors we choose $\beta =0.63$. The non-window optical depth has the same expression as the grey-atmosphere one,
\begin{equation}
\label{eq:taulw}
	\tau^{\lw} = \tau^{\lw}_{\ws}\exp(-z/H_a)+\tau^{\lw}_{\ds}\exp(-z/H_s).
\end{equation}
However, the values of surface optical depth must be increased to account for the fact that only a portion of the infra-red radiation goes through this band. For the window region the main absorbers are ozone (which we do not treat here) and water vapour \citep[e.g.,][Fig.\ 3.14]{Andrews10}, and thus we take the window optical depth to vary as
\begin{equation}
\label{eq:tauwin}
	\tau^{\win} = \tau^{\win}_{\ws}\exp(-z/H_a),
\end{equation}
The surface optical depth in the window region is quite small compared with that in the non-window region.

As in the grey-atmosphere configuration, we assume that the stratosphere is in radiative equilibrium, which now requires
\begin{equation}
\label{eq:dIzero2band}
     \frac {\partial (U^{\lw}+U^{\win}-D^{\lw}-D^{\win})} {\partial z} = 0.
\end{equation}
Since the stratospheric optical depth is extremely small in the window band, $U^{\win}$ and $D^{\win}$ are nearly constant with height. As a result, the primary balance in Eq.~(\ref{eq:dIzero2band}) is between $\partial U^{\lw}/\partial z$ and $\partial D^{\lw}/\partial z$. In other words, the non-window band is in near \RE in the stratosphere, namely,
\begin{equation}
\label{eq:dIzero2bandapprox}
     \frac {\partial (U^{\lw}-D^{\lw})} {\partial z} \approx 0.
\end{equation}
The system is closed by the \TOA boundary conditions: $D^{\lw} = 0$, $D^{\win}=0$, $U^{\lw} = \OLRT^{\lw}$ and $U^{\win} = \OLRT^{\win}$, where $\OLRT^{\lw}+\OLRT^{\win}= \OLRT$. In addition, we have the surface boundary conditions: $U^{\lw} = \beta \sigma T_s^4$ and $U^{\win} = (1-\beta) \sigma T_s^4$ where $T_s$ is surface temperature.

There are similarities between the non-window band and the grey-atmosphere model, specifically in the equivalence of Eqs.~(\ref{eq:dIzero2bandapprox}) and  (\ref{eq:dIzero}). In the stratosphere, where $\tau^{\win}$ is negligible, the  \RE temperature in the two-band model is given by 
\begin{equation}
\label{eq:Tst2band}
     T_{\re} = \left[ \left(\frac{\tau^{\lw}+1}{2\sigma}\right)\OLRT^{\lw} \right]^{\tfrac{1}{4}}.
\end{equation}
Of course it is OLR that is a boundary condition in the model, not OLR$^{\lw}$, but (\ref{eq:Tst2band}) will be a useful relation as discussed below.  Eq. (\ref{eq:Tst2band}) is very similar to Eq.~(\ref{eq:Tst}), the equivalent expression in the grey model. 

\begin{figure} [t!]
\centering{\includegraphics[width=0.9\textwidth]{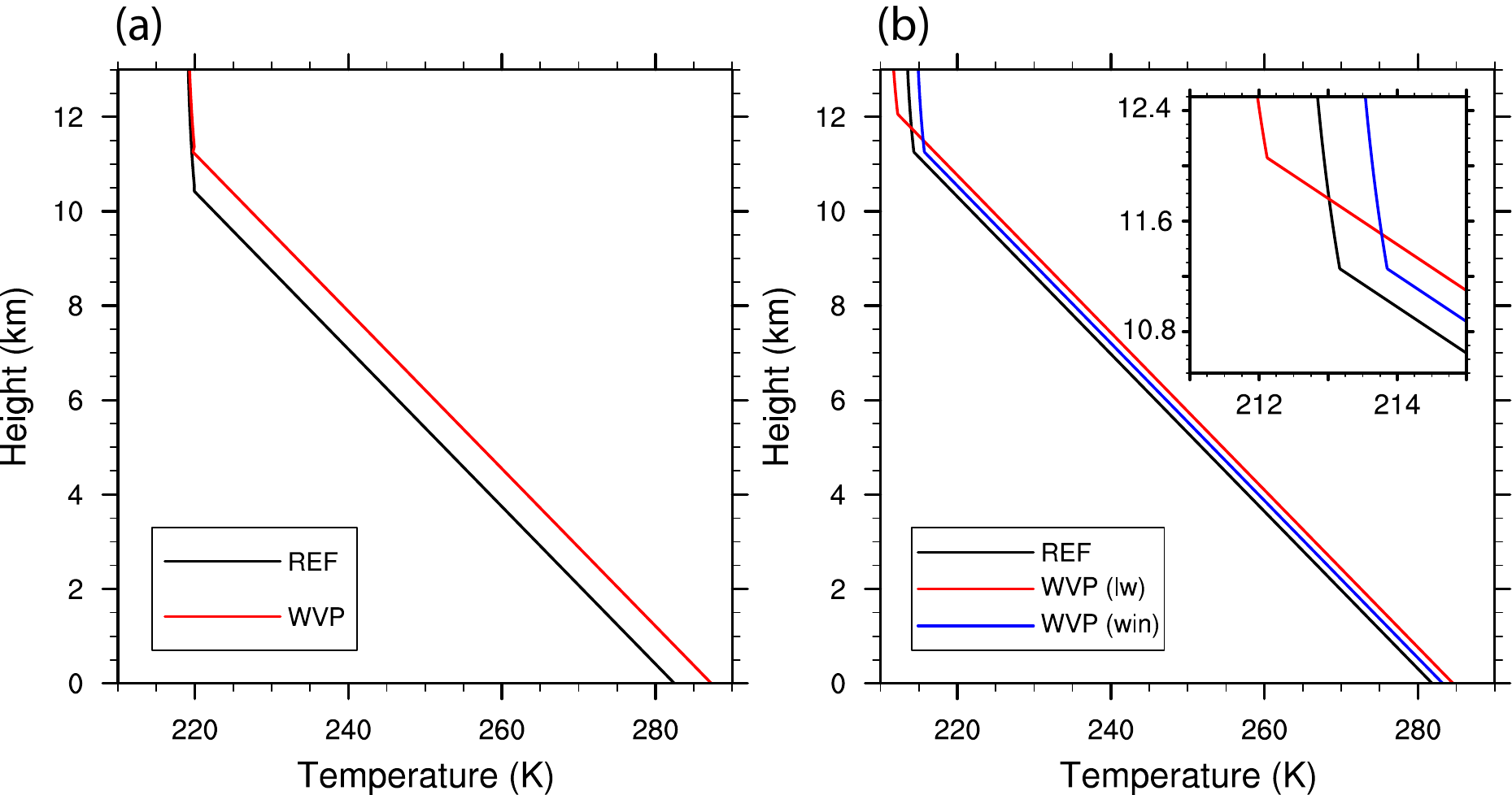}}
\caption{Temperature profile response to \WVP increase (a) in the grey model and (b) in the two-band model. In the reference case (REF), we set OLR = 260 W/m$^2$, the \TLR $\Gamma$ = 6 K/km, the surface optical depth $\tau_{\ws}=4$ for the grey model, $\tau^{\lw}_{\ws}=8$ and $\tau^{\win}_{\ws}=1$ for the two-band model. For each WVP case, we assume that the surface optical depth increases by 50\%.}
\label{WVP2band}
\end{figure}

In the grey model, an increase of $\tau$ in the troposphere leads to very little change in tropopause temperature, because the tropopause is simply extending  into a nearly isothermal lower stratosphere.  However, in a windowed model the response depends on whether the increase in optical path occurs in the window or non-window region, as we now illustrate with some idealized calculations (Fig.~\ref{WVP2band}). We assume that water vapor is the only infra-red absorber in the atmosphere in both window and non-window regions ($\tau_{\ds}=0$, $\tau^{\lw}_{\ds}=0$). We set OLR = 260 W/m$^2$ and \TLR $\Gamma$ = 6 K/km. We set the surface optical depth $\tau_{\ws}=4$ for the grey model, $\tau^{\lw}_{\ws}=8$ and  $\tau^{\win}_{\ws}=1$ for the two-band model. We assume that the surface optical depth increases by 50\% under global warming and investigate the tropopause changes in three cases: (i) the grey model, and, for the two-band model, (ii) an increase in optical depth in the non-window region only, and  (iii) an increase in the window region only.  

As noted, in the grey model the tropopause temperature stays virtually constant (Fig.~\ref{WVP2band}a).
In case (ii) we incorporate the \WVP increase solely as an increase in $\tau^{\lw}_{\ws}$, while keeping the optical depth in the window band unchanged. In response to the $\tau^{\lw}_{\ws}$ increase, the surface temperature increases, OLR$^{\win}$ increases (because the increased infra-red radiation emitted from the ground passes upwards virtually unobstructed), and therefore OLR$^{\lw}$ must decrease to keep total OLR the same. For a 50\% increase of $\tau^{\lw}_{\ws}$, we find that the equilibrium $\Delta$OLR$^{\lw}=-$OLR$^{\win}=-4.8$ W/m$^2$. From Eq.~(\ref{eq:Tst2band}) the stratospheric temperature falls. A warmer troposphere and a cooler stratosphere implies that they have to meet at a higher level (i.e., there is an increase of tropopause height) and at a colder temperature (i.e. a decrease of tropopause temperature), and the results of the calculation are shown in red line in Fig.~\ref{WVP2band}b.
 
In the other limit, case (iii), we incorporate the \WVP increase as an increase in $\tau^{\win}_{\ws}$,  keeping the optical depth in the non-window band unchanged.  The possibility that this can lead to an increase in tropopause temperature can be revealed by a simple argument.  The \OLR in the window band nearly all comes from the surface, and is equal to $(1-\beta) \textit{Tr} \,\sigma T_s^4$ where \textit{Tr} is the transmissivity of the atmosphere. Now $T_s$ increases as \WVP increases, for this is the greenhouse effect. However, \textit{Tr} decreases, because there is more atmospheric absorption. If the \textit{Tr} decrease is sufficiently large there will be\textit{ less} radiation to space from the ground in the window region. The radiation in the non-window region must compensate for this, and from Eq.~(\ref{eq:Tst2band}) the tropopause temperature must then increase if OLR is to stay the same.  A calculation with the two-band model illustrates this effect more quantitatively --- see the blue line in Fig.~\ref{WVP2band}b. In response to a 50\% increase in $\tau^{\win}_{\ws}$,  the surface temperature and tropospheric temperature both increase, but the tropopause height does not change very much  (Fig.~\ref{WVP2band}b).  Then, given the unchanged optical depth profile in the non-window band, there is an increase in OLR$^{\lw}$ (+3.2 W/m$^2$ for this case), compensating for the decrease in OLR$^{\win}$ and keeping the total OLR the same.

\begin{figure} [t]
\centering{\includegraphics[width=0.9\textwidth]{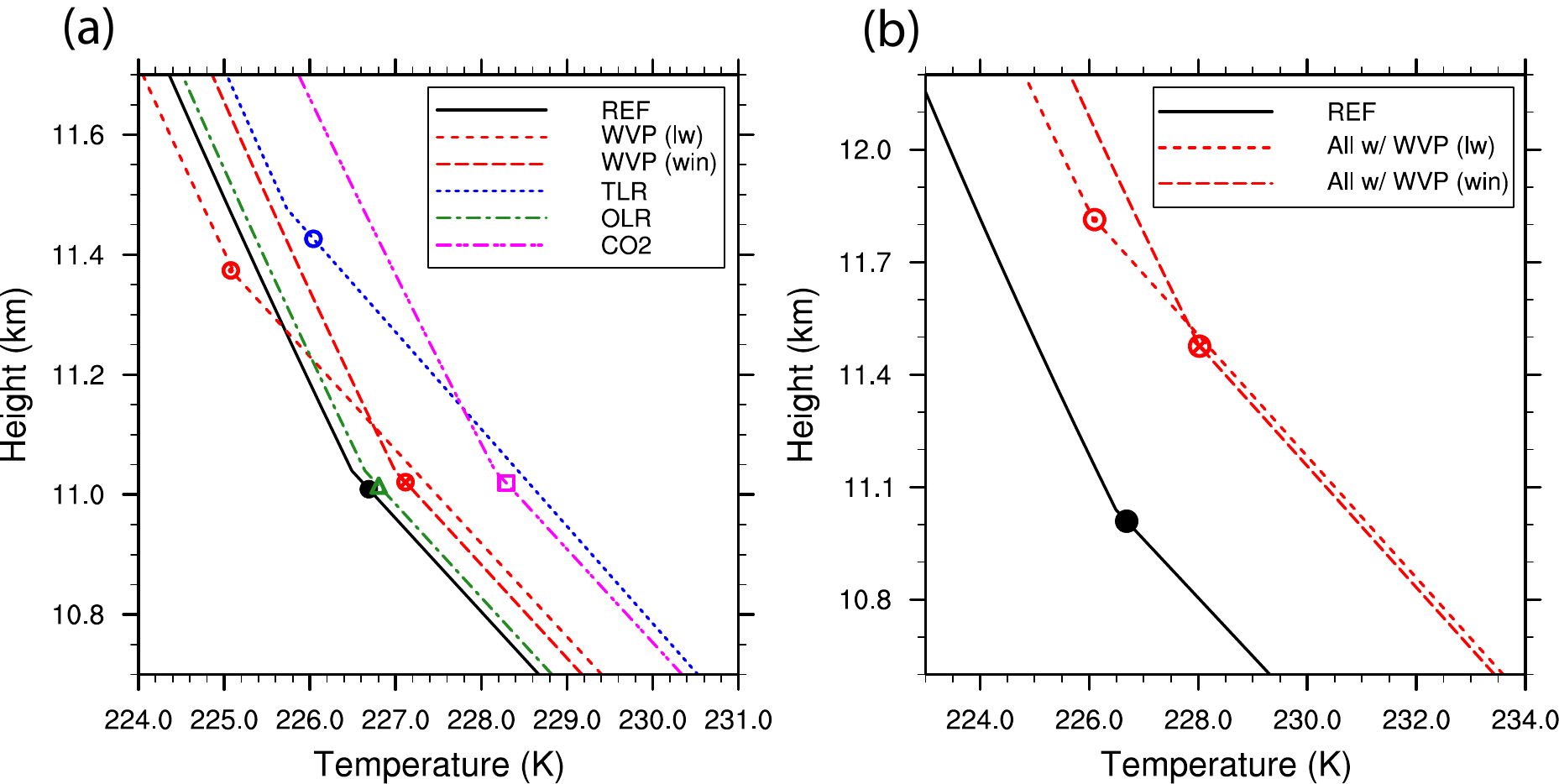}}
\caption{(a) Temperature profiles and point of tropical tropopause produced by the two-band tropopause model. `NCEP2' refers to the reference state using tropical mean (30\textdegree S-30\textdegree N) NCEP2-informed input variables. `WVP (lw)' and `WVP (win)' refers to the cases where CMIP5 \WVP changes are incorporated as the optical depth increase in the non-window and window regions, respectively, while other input variables keep unchanged. `TLR', `OLR' and `CO2' refers to the cases with the CMIP5 changes in \TLR, \OLR and carbon dioxide, respectively, alone. (b) Similar to panel a but for the cases with all the four effects (WVP, TLR, OLR and CO2) with the two distinctive treatments of WVP increase. For all the calculations, we first apply the tropical-mean multi-model-mean changes in input variables and then identify the tropopause (thicker dots) here defined as in Eq.~(\ref{eq:Tz}) as the top of the region of imposed stratification.
}
\label{Trop2band}
\end{figure}

In actuality, both the window and non-window optical depths may increase, potentially leading to increases in both tropopause height and temperature, as is seen in the CMIP5 models.  To illustrate this we set the model parameters to more realistic values, and in particular, now we add back the contribution of carbon dioxide to the non-window optical depth. We set $\tau^{\lw}_{\ds} = 2.6$, and $\tau^{\lw}_{\ws} = \alpha^{\lw} \text{WVP}$, where $\alpha^{\lw}=0.26$ mm$^{-1}$. We set $\tau^{\win}_{\ws} = \alpha^{\win} \text{WVP}$, where $\alpha^{\win}=0.0125$ mm$^{-1}$. In the tropics, \WVP is in the range of 20--40 mm (Fig.~\ref{input}a), and it can be translated into a $\tau^{\win}_{\ws}$ range of about 0.25--0.5. It corresponds to a range of absorption percentage of about 20-40\%. (Note that the infra-red absorption and emission in the window band are more important in the tropics than elsewhere \citep[e.g.,][]{huang2014} because of the larger amount of water vapour.) The resultant mean climate generally resembles that from the grey-atmosphere model, and has a \GMST of 15\textdegree C. 

We then use the model parameters given above and impose the changes in input variables (WVP, TLR, OLR, and CO$_2$) from the CMIP5 models to the NCEP2 basic state. For the first three input variables, they are first averaged across the various GCMs and across the latitudinal bands in the tropics (30\textdegree S-30\textdegree N), and then applied to the two-band model. For \COT, an equivalent $\tau^{\lw}_{\ds}$ increase of 0.43 is imposed, which would roughly give rise to a radiative forcing of an increase of CO$_2$ by a factor of 2.7, which is about 100 years of 1\%/year increase).  We perform experiments with the water vapour optical path increased only in the non-window region, only in the window region, and in both. 

The changes in tropopause height and temperature found in this more realistic configuration have the same characteristics as those found in the idealized calculations. In the case with $\tau^{\lw}_{\ws}$ alone increasing, the \WVP increase alone would lead to a decrease of tropical tropopause temperature by -1.6\textdegree C per century (Fig.~\ref{Trop2band}a). In the case with $\tau^{\win}_{\ws}$ alone increasing, the same \WVP increase would result in an increase of tropopause temperature by +0.4\textdegree C per century. After accounting for all the four factors (CO$_2$, WVP, TLR and OLR), the model predicts a tropical tropopause temperature change of -0.6\textdegree C for the non-window case, and +1.4\textdegree C per century for the window case (Fig.~\ref{Trop2band}b). The calculation in which the increase in optical depth is split across window and non-window regions (not shown) lies between these two cases.

The results from the case in which the optical path increases in the window region agree quite well with the CMIP5 model projections (about 1.3\textdegree C per century). However, the spread in the tropopause temperature changes among the CMIP5 models is not well captured by the two-band model (not shown). The real atmosphere has the contribution from both effects, but the effect associated with the window case most likely dominates, in particular over the tropics \citep{huang2014}, because the infra-red absorption in the non-window band is nearly saturated and the overlap of absorption between CO$_2$ and water vapour is particularly large.

\subsection{Tropopause height changes in the windowed model}  \label{sec:twoband}

\begin{figure} [t]
\centering
\includegraphics[width=0.9\textwidth]{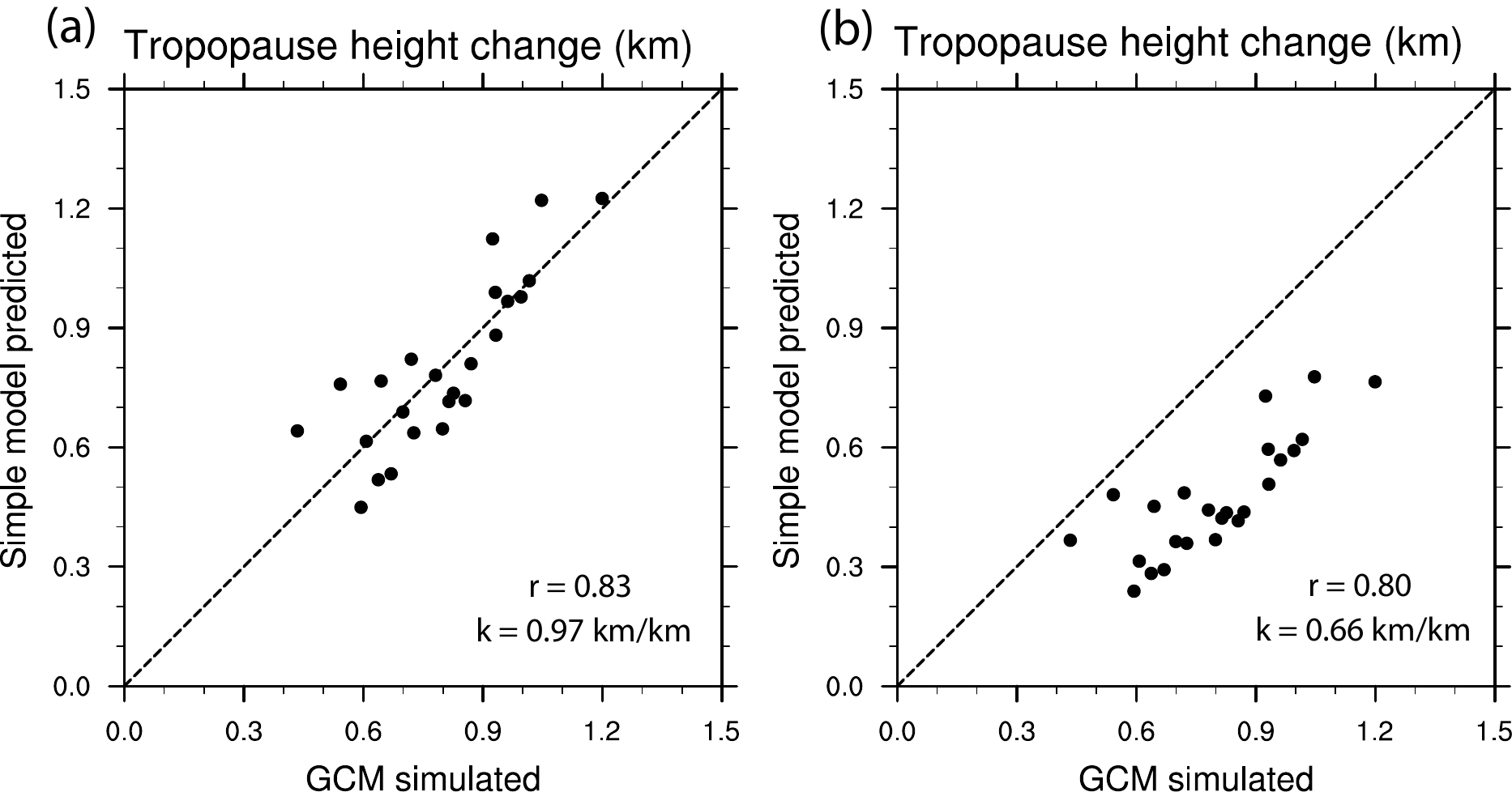}  
\caption{As for fig.\ \ref{pred_trop_change}, showing scatter plots of model-predicted versus GCM simulations of tropopause height, but now using the two band model. (a) Optical depth increase in the non-windowed region only. (b) Optical depth increase in the windowed region only.}
\label{height2band}
\end{figure}

We see in \figref{Trop2band} that if the increase in optical depth is in the window region then the tropopause height increase is smaller than if the optical depth increase is in the non-window region, which in turn is similar to the  grey-model predictions. To explore this further, we repeat the calculations described in Section \ref{sec:TT} with the two-band model, and results analogous to those of \figref{pred_trop_change} are shown in \figref{height2band}. If the optical path increase is applied solely to the non-window region, as in panel (a), then the results are very similar to those of the grey model. If the optical path increase is applied in the window region then the tropopause height increase is systematically smaller. This result is expected because changes in the optical path in the window region has only a small direct effect on changes in tropopause height, and the increase in height is mostly due to changes in lapse rate; the height change is therefore smaller than when the optical path increases in the non-window region. There is, nevertheless, still a very good correlation between the simple model and the GCM results in all cases.  In reality, an increase in water vapour content will affect both the window and non-window parts of the spectrum, in different proportions depending on water vapour content and cloudiness, leading to an increase in both tropopause height and temperature. A quantitative treatment of all these effects would require not only a full multi-band radiation model but also a good model of clouds, which is beyond our scope. 
 
It is evident that the grey model captures the future increases in tropopause height, with some quantitative changes then arising from the non-greyness of the atmosphere.  However, changes in tropopause temperature, although small compared to changes in surface temperature, require a windowed radiative model to explain them. (The grey model does explain the lowest-order result that tropopause temperature changes are small compared to surface temperature changes.) There are of course other potential mechanisms that might affect the tropopause height and temperature that occur in GCMs but are not considered in our study, such as changes dynamical cooling induced by the stratospheric circulation because of changes in the Brewer--Dobson circulation or ozone heating.  

\section{Summary and Conclusions}  
\label{sec:conclusions}

In this study we have used a relatively simple column model to predict the meridional structure and possible future changes of height and temperature of the tropopause associated with global warming. The model assumes a troposphere with a given lapse rate that connects continuously to a stratosphere in which the lapse rate is determined by either  pure radiative balance or a radiative-dynamical balance.  If we assume that the tropospheric lapse rate is determined by convection then the model is essentially a radiative--convective model.  The model explicitly exposes the dependence of tropopause height on the optical depth, \TLR, and \OLR of the atmosphere. Thus, a greater optical depth or a smaller \TLR will elevate the tropopause, whereas the tropopause height is relatively insensitive to the \OLR changes, as may be inferred from the approximate analytic solution Eq.~(\ref{trop_analytic}). 

When applied to the present climate, the model, even when configured with a stratosphere in radiative equilibrium, is able to reproduce the meridional shape of the tropopause height --- one that is higher in the tropics and lower in the polar regions with a fairly sharp transition in the extratropics, as in \figref{pred_trop}. The higher tropopause in the tropics here results from the greater \WVP,  but its impact is largely compensated by that of the larger tropical \TLR which reduces tropopause height. As a result, the equator-to-pole contrast in tropopause height predicted by the model is too small compared with the observations.  If we incorporate a dynamical cooling profile to represent the impact of the stratospheric circulation then the tropical tropopause is elevated and cooled, making it closer to what is observed, broadly consistent with the results of \citet{thuburn2000} and \citet{haqq2011}.  In this case, the tropical tropopause, as defined by a lapse-rate criterion,  is well above the boundary between a convective region (with a given lapse rate) and a region where the lapse rate is determined by a balance between radiation and slow dynamics.  A thermal (or WMO) tropopause is then not a particularly good demarcation between the dynamics of the troposphere and stratosphere. 

The model may be used to disentangle the multiple factors affecting the change in tropopause height and temperature associated with global warming.  An increase in tropopause height is one of the most robust consequences of such warming, predicted by nearly all CMIP5 models in nearly all warming scenarios. In the 1\% scenario (with CO$_2$ increasing by 1\% per year) the annual-mean zonal-mean tropopause height increases, on average, by about 0.7 km within a century for all the latitudes from the equator to the poles, although there is a large spread of the increase among the models. In lower latitudes (50\textdegree S-50\textdegree N) the inter-model spread is highly correlated with, and most likely due to, the differences in model-predicted changes in \WVP and \TLR, while in the high-latitudes significant correlations are only found with the \WVP changes in the Northern Hemisphere. 

Globally averaged, the grey model predicts the magnitude of tropical-mean tropopause height increase for each individual CMIP5 model quite well, after incorporating the effects of CO$_2$, \WVP, \TLR and \OLR. (The inter-model correlation between the simple-model-predicted and the GCM-simulated tropopause height changes is above 0.8, and the regression coefficient is close to one.) Among the four controlling factors, the contribution of changes in \WVP and lapse rate dominate, with the direct effects of changes  CO$_2$ and \OLR lagging far behind. Even though CO$_2$ is the only external forcing, its direct contribution on the tropopause height increase is small. The presence of infra-red window will make the \WVP effect a little less important, but still comparable to the lapse rate effect.

The tropopause temperature, as well as height, is found to increase in CMIP models with global warming, especially in low latitudes, although the average increase is much less than the surface temperature increase.  Now,  if the radiative transfer is grey in the infra-red and if the stratosphere is in radiative equilibrium with a small optical depth then the tropopause temperature will stay the same with global warming. That is, the `fixed tropopause temperature' hypothesis is exact in these circumstances. The small increase in tropopause temperature in more complex models and observations may, however, be explained by the presence of  infra-red window, if an increase in optical depth occurs in the window region. In this circumstance the \OLR in the window region decreases and must be compensated by an increase in \OLR from the non-window region, entailing an increase in tropopause temperature. The fact that temperature increases are greater at low latitudes, where water vapour effects are strongest, is consistent with this hypothesis.  In this paper we do not speak to the issue of whether and how changes in tropopause temperature are related to, or even violate, the fixed-anvil-temperature hypothesis.

The model we have presented and used has a number of limitations, and in particular it has limited skill in high-latitudes. This is likely because a vertically-uniform \TLR is not a good assumption in the presence of an inversion layer in the lower troposphere. (The model can easily be extended to have a non-uniform lapse rate but this would be a little arbitrary.) The model also makes a number of other assumptions and is (deliberately) not complete --- the lapse rate and \WVP are specified and not part of the model solution, for example.  Thus, although the simple model enables an attribution to be made of why and how the various CMIP models differ in their responses, it does not explain why those differences (for example, in water vapour content) arise, nor does it seek to explain any effects arising from changes in meridional heat transfer. These topics merit future investigation.


\section*{acknowledgements}
S. Hu is supported by the Scripps Institutional Postdoctoral Program Fellowship. G. K. Vallis is supported by the Leverhulme Trust and NERC. This work was begun at the Geophysical Fluid Dynamics program at Woods Hole Oceanographic Institution and we acknowledge all present for a stimulating atmosphere.


\section*{Appendix}
\subsection*{Numerical solution of the tropopause model}
\renewcommand{\theequation}{A\arabic{equation}}
\setcounter{equation}{0}
In this section we provide more details on how the height of the tropopause is numerically solved. We will start with the case that has a specified stratospheric dynamical heating profile $Q_s$ and then discuss the special case with a \RE stratosphere (i.e. $Q_s = 0$) that is used for the main body of our study. For clarity of the discussion, in a few places we repeat some of the equations that are already shown in the main text.

We assume a grey atmosphere that is transparent to the solar radiation and write the infra-red longwave radiative transfer equations as
\begin{equation}
\label{Aeq:dDdU}
     \frac {\partial D} {\partial \tau} = B - D, \qquad \frac {\partial U} {\partial \tau} = U - B.
\end{equation}
$D$ and $U$ are downward and upward infra-red irradiance, respectively, $B=\sigma T^4$ follows the Stefan-Boltzmann law, with $\sigma=5.67x10^{-8}$~W~m$^{-2}$, and $\tau$ is optical depth increasing downward. The upper boundary conditions (at the top of the atmosphere) are that $D=0$ and $U= \OLRT$ at $\tau=0$, and we take OLR to be given. 

For the ease of calculation, we define two variables, $I$ and $J$, using the linear combinations of $U$ and $D$, as follows,
\begin{equation}
\label{Aeq:IJ}
     I = U - D, \qquad J = U + D.
\end{equation}
As a result, Eq.~(\ref{Aeq:dDdU}) can be written as, equivalently,
\begin{equation}
\label{Aeq:dIdJ}
     \frac {\partial I} {\partial \tau} = J - 2B, \qquad \frac {\partial J} {\partial \tau} = I.
\end{equation}
The upper boundary conditions thus become: $I= \OLRT$ and $J= \OLRT$ at $\tau=0$. We assume \RDE in the upper atmosphere, that is, the net convergence of the infra-red radiation is balanced by the dynamical cooling induced by the stratospheric circulation,
\begin{equation}
\label{Aeq:dIQs}
     \frac {\partial I} {\partial \tau}+Q_s = 0.
\end{equation}
Using Eq.~(\ref{Aeq:dIQs}) and the upper boundary condition, we compute the vertical profile of $I$,
\begin{equation}
\label{Aeq:Isoln}
     I = \OLRT - \overline{Q_s},
\end{equation}
where $\overline{Q_s}(\tau)=\int_{0}^{\tau} Q_s d\tau'$. Combining Eq.~(\ref{Aeq:Isoln}) with Eq.~(\ref{Aeq:dIdJ}), we get the vertical profile of $J$,
\begin{equation}
\label{Aeq:Jsoln}
     J = (\tau+1)\, \OLRT - \overline{\overline{Q_s}},
\end{equation}
and the vertical profile of $B$,
\begin{equation}
\label{Aeq:Bsoln}
     B = \left(\frac{\tau+1}{2}\right)\OLRT + \frac{Q_s - \overline{\overline{Q_s}}}{2},
\end{equation}
where $\overline{\overline{Q_s}}(\tau)=\int_{0}^{\tau} \overline{Q_s} d\tau'$. Using the Stefan-Boltzman law, we get the \RDE temperature profile
\begin{equation}
\label{Aeq:Tst}
     T_{\rde} = \left[ \left(\frac{\tau+1}{2\sigma}\right)\OLRT + \frac{Q_s - \overline{\overline{Q_s}}}{2\sigma} \right]^{\frac{1}{4}}.
\end{equation}
Note that the \RDE temperature profile is derived without any lower boundary conditions at the surface. Combining Eqs. (\ref{Aeq:IJ}), (\ref{Aeq:Isoln}), (\ref{Aeq:Jsoln}), we have the vertical profiles of $U$ and $D$,
\begin{equation}
\label{Aeq:UDsoln}
     U = \left(\frac{\tau+2}{2}\right)\OLRT - \frac{\overline{Q_s} + \overline{\overline{Q_s}}}{2}, \qquad D = \left(\frac{\tau}{2}\right)\OLRT + \frac{\overline{Q_s} - \overline{\overline{Q_s}}}{2}.
\end{equation}

In this study, we prescribe the vertical profile of optical depth $\tau = \tau(z)$, which may consist of one or two infra-red absorbers. We assume the exponential decrease of pressure with height $p=p_s\exp(-z/H_s)$, where surface pressure $p_s = 1000$ hPa and the scale height of dry air $H_s = 8$ km. Therefore, one can convert the vertical coordinates of $\tau$, $p$ and $z$, from one to another.

We assume that the lower atmosphere below the \RDE layer, separated by the boundary at $z=H_T$, is uniformly stratified with a specified lapse rate $\Gamma$, namely, 
\begin{equation}
\label{Aeq:Tz}
	T(z) = \begin{cases}
	T_{\rde}(z), & z \geq H_T, \\
	T_T+\Gamma(H_T-z), & H_T \geq z \geq 0,
	\end{cases}
\end{equation}
where $T_T=T_{\rde}|_{z=H_T}$. The lower boundary condition at the surface requires that $U=\sigma T_s^4$ at $z=0$, where $T_s$ is the surface temperature (no ground temperature jump). To summarize, with the specified $\OLRT$, $\Gamma$, $\tau(z)$ and $Q_s(z)$, the only unknown variable in the system is the tropopause height $H_T$. Therefore, we can numerically solve the system by iterating over the different values of $H_T$ until the lower boundary condition is matched. Note that the numerical solution of $H_T$ should be literally interpreted as the boundary between the upper \RDE layer and the lower uniformly stratified layer, but not the tropopause height; see Section \ref{sec:stratosphere} for an illustrative example.

In the main body of this study, our tropopause model does not involve the stratospheric dynamical heating and, in other words, $Q_s$ vanishes. In that case, we have the \RE solutions instead for the upper atmosphere, that is,
\begin{equation}
\label{eq:DUBsoln_Qs0}
     D,U,B = \left( \frac{\tau}{2}, \frac{\tau+2}{2}, \frac{\tau+1}{2} \right) \OLRT,
\end{equation}
and
\begin{equation}
\label{Aeq:Tsoln_Qs0}
     T_{\re} = \left[ \left(\frac{\tau+1}{2\sigma}\right) \OLRT \right]^{\frac{1}{4}}.
\end{equation}

In our two-band model, we have two downward and two upward infra-red irradiances ($D^{\lw}$, $D^{\win}$, $U^{\lw}$ and $U^{\win}$) instead of one each as in our grey model ($D$ and $U$). We also have five boundary conditions ($D^{\lw}=0$, $D^{\win}=0$ and $U^{\lw}+U^{\win}=\OLRT$ at the top of atmosphere, and $U^{\lw}=\beta \sigma T_s^4$ and $U^{\win}=(1-\beta) \sigma T_s^4$ at the surface) instead of three ($D=0$ and $U=\OLRT$ at the top of atmosphere, and $U=\sigma T_s^4$ at the surface). Thus, the system is still closed and the height of tropopause can be numerically solved using an iterative method similar to that used in the grey scheme.

\setlength{\bibsep}{5pt}
\bibliography{HV19}

\end{document}